\documentclass[aps,showpacs]{revtex4}

\usepackage{graphicx}
\usepackage{epsfig}
\newcommand{\ffat}[1]{\mbox {\boldmath $#1$}}
\newcommand{\hatbf}[1]{\hat{{\bf #1}}}

\begin{document}
\bibliographystyle{apsrev}
%\draft

\title{The Operator form of $^3$H ($^3$He) and its Spin Structure}

\author{I.~Fachruddin\thanks{Permanent address: Jurusan Fisika, 
Universitas Indonesia, Depok 16424, Indonesia}, W. Gl\"ockle} 
\affiliation{Institut f\"ur Theoretische Physik II, Ruhr-Universit\"at Bochum, 
D-44780 Bochum, Germany.}

\author{Ch. Elster}
\affiliation{Department of Physics and Astronomy, Ohio University, Athens, OH 34701, USA}
%\\and Institut f\"ur Kernphysik, Forschungszentrum J\"ulich, D-52425 J\"ulich,
%Germany}

\author{A. Nogga}
\affiliation{
Institute for Nuclear Theory, University of Washington, Seattle, WA 98195, USA  }

\vspace{10mm}

\date{\today}

%\maketitle

\begin{abstract}
 An operator form of the 3N bound state is proposed. It consists of
eight operators formed out of scalar products in relative momentum and
spin vectors, which are applied on a pure 3N spin 1/2 state. Each of the 
operators is associated with a scalar function depending only on the magnitudes
of the two relative momenta and the angle between them. The connection
between the standard partial wave decomposition of the 3N bound state and
the operator form is established, and the decomposition of
these scalar function  in terms of partial wave components and
analytically known auxiliary functions is given. That newly established operator
form of the 3N bound state exhibits the dominant angular and
spin dependence analytically. The scalar functions are tabulated
and can be downloaded. As an application the spin dependent nucleon momentum
distribution in a polarized 3N bound state is calculated to illustrate the use of
the new form of the 3N bound state.
\end{abstract}

\vspace{10mm}


\maketitle
%\pagebreak

%****************************************************************************

%\narrowtext

%****************************************************************************

\section{Introduction}

Today it can be considered as standard to solve the Schr\"odinger equation for three
nucleons numerically with high precision. This can be done either in the form of
Faddeev equations \cite{nogga00,nogga02b}, using hyperspherical expansion \cite{kievsky93,kievsky97} 
or the Gaussian-basis method \cite{kamimura88}. Also the Greens-Function-Monte-Carlo (GFMC) 
and No-Core-Shell Model approaches have been applied to the three nucleon system 
\cite{pieper01b,navratil98a}.
These methods except for GFMC have always been based on standard partial wave
expansions. Only recently for the study of three bound bosons the Faddeev equation has
been solved directly in terms of relative momentum vectors
\cite{elster99,liu02,liu03}. Though any expectation value for a three-nucleon (3N) bound
state can be determined numerically from wave functions obtained by any of the above
mentioned methods, no analytic insight can be extracted for the spin and momentum (or
configuration) dependence of the 3N bound state.

For the deuteron Rarita and Schwinger \cite{rarita41a,rarita41b} introduced an operator
form of the deuteron state in terms of spin operators and the relative position
vector. This representation is ideal to exhibit the probabilities for finding any
spin orientations in relation with the relative position (or momentum) vectors of the
two nucleon in a polarized deuteron. In the case of a momentum representation of the
deuteron derived from modern nucleon-nucleon (NN) forces this is illustrated 
in Ref. \cite{fachruddin01} and
for a coordinate representation in Ref. \cite{forest96}.

For three nucleons it is much more difficult to express the spin and momentum
dependence of the bound state in an analytic form. Again, this analytic form has been
worked out long time ago by Gerjuoy and Schwinger \cite{gerjuoy42}. For the 3N bound
state, however, the functions multiplying the different scalars built from spin and
position vectors depend on three variables, namely the magnitudes of the Jacobi
vectors and the angle between them. In Ref.~\cite{gerjuoy42} the wave 
function is expanded in nine operators with nine corresponding  
scalar functions, which were numerically unknown in those days. 
It is the aim of this
paper to establish an analytic link between these functions 
and the usual partial wave representation of the 3N bound state. This will 
lead to an alternative representation of the $^3$H ($^3$He) bound states, which 
is more accessible to analytic insights into the spin structure of the 
3N bound state.
In our analysis we also find that the last term given in 
Ref. \cite{gerjuoy42} is redundant, further simplifying this representation.

The paper is organized as follows. In Section II we start from the
standard partial wave representation of the  3N bound state and reformulate it in
terms of scalar operators acting on pure spin states with the correct spin quantum
numbers of $^3$H ($^3$He). This reformulation then leads at the same time to the scalar functions.
Section III is devoted to the numerical investigation of the scalar functions. In Section IV we give
an example for the application of the operator form of the $^3$He wave function and evaluate the
probability to find a neutron with given momentum vector and polarized in the direction of
the overall polarization of $^3$He. Several appendices include technical steps for the derivations.
Finally we summarize in Section V.

\section{The Operator Form of the $^3$H ($^3$He) Bound State}
\subsection{Derivation}

The starting point for the derivation of an operator form is the standard 
partial wave representation of a 3N bound state. Here we do not use the isospin
formalism and choose particles $1$ and $2$ to be neutrons(protons) and particle $3$ to
be the proton(neutron). For the derivation we will assume a $^3$H bound state, but 
the generalization to $^3$He is obvious. 
In momentum space the 3N bound state with total angular momentum $1/2$ and
magnetic quantum number $m$ can be written as
\begin{eqnarray}
\Psi^m({\bf p}, {\bf q}) &\equiv & \langle {\bf p} {\bf q} | \Psi^m \rangle \nonumber \\
 &=& \sum_{sSM_S} |(s \frac{1}{2}) S M_S \rangle \langle (s \frac{1}{2}) S M_S |
                   \langle {\bf p} {\bf q} | \Psi^m \rangle \nonumber \\
&=& \sum_{sSM_S} |(s \frac{1}{2}) S M_S \rangle \sum_{l \lambda L} 
   C(LS\frac{1}{2}, m-M_S \: M_S) {\mathcal Y}^{L m-M_S}_{l \lambda}({\hat {\bf p}} {\hat {\bf q}})
                            \:   \Psi_{l \lambda L s S}(p q) \nonumber \\
&=& \sum_{sS} \sum_{l \lambda L} \left[ {\mathcal Y}^L_{l \lambda} ({\hat {\bf p}} {\hat {\bf q}}) 
  |(s \frac{1}{2}) S \rangle \right] ^{\frac{1}{2} m} \: \Psi_{l \lambda L s S}(p q) ,
\label{eq:2.1}
\end{eqnarray}
where ${\bf p}$ and ${\bf q}$ are the standard Jacobi momenta
\cite{glocklefb}, and the ${\hat {\bf p}}$ and ${\hat {\bf q}}$ stand for the
corresponding unit vectors.
The 3N spin state $ |(s \frac{1}{2}) S M_S \rangle$ is constructed by coupling the
spin state of the two neutrons with total spin $s$ and the spin of the proton 
$(\frac{1}{2})$ to the total spin $S$ and magnetic quantum number $M_S$ of the 
3N system. Furthermore, $l$, $\lambda$, and $L$ are the relative orbital angular
momenta of the two neutrons (related to ${\bf p}$),  the orbital angular momentum of
the proton (related to ${\bf q}$) and the total orbital angular momentum of the
three nucleons. The  bracket in the last equation is a convenient abbreviation
for the LS-coupling employed here. The quantities $\Psi_{l \lambda L s S}(p q)$
represent the partial wave components of the 3N bound state. They are for example
determined by the solution of the Faddeev equations \cite{nogga00,nogga02b}. Typical numbers for a good
representation of $\Psi^m({\bf p}, {\bf q})$ are $l\leq 3$, $\lambda \leq 3$. 
The wave function is
antisymmetric in the two neutrons, which constrains $l+s$ to be even. 
Note that we solve the Schr\"odinger equation using the isospin formalism. 
The particle basis wave functions, which enter in Eq.~(\ref{eq:2.1}), 
are a combination of total isospin $T=1/2$ and $T=3/2$ 
wave functions, namely
\begin{equation}
\label{eq:iso}
\Psi_{l \lambda L s S}(p q) = \sqrt{3} \ \left( \ (-)^{M_T+3/2} \ \sqrt{2 \over 3} \  
\Psi_{l \lambda L s S}^{t=1\ T={1 \over 2}}(p q)  +  \sqrt{1 \over 3} \  
\Psi_{l \lambda L s S}^{t=1\ T={3 \over 2}}(p q)   \right) 
\end{equation}
where $M_T$ is the third component of the isospin, namely $M_T=1/2$ for $^3$He and $-1/2$ for 
$^3$H. Note that the isospin of the neutron-neutron or proton-proton  two-body 
subsystem is restricted to $t=1$.
The overall factor $\sqrt{3}$ is a consequence of the identity of the nucleons 
in the isospin formalism and insures the correct normalization. 

The right hand side of Eq.~(\ref{eq:2.1}) naturally decomposes into four parts
according to the total spin $S=1/2$ or $S=3/2$ and the total orbital angular momentum
$L=0,1$, and $2$ as
\begin{eqnarray}
\Psi^m({\bf p}, {\bf q}) &=& \sum_s \sum_l \left[ {\mathcal Y}^0_{ll}({\hat {\bf p}}
   {\hat {\bf q}}) \mid (s \frac{1}{2}) \frac{1}{2} \rangle \right] ^{\frac{1}{2} m} \:
\Psi_{ll0s \frac{1}{2}}(p q) \nonumber \\
& & + \sum_s \sum_{l \lambda} \left[ {\mathcal Y}^1_{l\lambda} ({\hat {\bf p}} {\hat {\bf q}})
\mid (s \frac{1}{2}) \frac{1}{2} \rangle \right] ^{\frac{1}{2} m} \: \Psi_{l
\lambda 1 s \frac{1}{2}}(p q) \nonumber \\
& & + \sum_{l \lambda} \left[ {\mathcal Y}^1_{l\lambda} ({\hat {\bf p}} {\hat {\bf q}}) \mid (1
\frac{1}{2}) \frac{3}{2} \rangle \right] ^{\frac{1}{2} m} \: \Psi_{l
\lambda 1 1 \frac{3}{2}}(p q)  \nonumber \\
& & + \sum_{l \lambda} \left[ {\mathcal Y}^2_{l\lambda} ({\hat {\bf p}} {\hat {\bf q}}) 
  \mid (1 \frac{1}{2}) \frac{3}{2} \rangle \right] ^{\frac{1}{2} m} \: \Psi_{l
\lambda 2 1 \frac{3}{2}}(p q)  \nonumber \\
& \equiv & ^2 S _{1/2} + ^2 P _{1/2} + ^4 P _{1/2} + ^4 D _{1/2}.
\label{eq:2.2}
\end{eqnarray} 
At first we consider the $^2 S _{1/2}$ part which can be separated in even and odd
terms with respect to $l$ as
\begin{eqnarray}
\mid \, ^2 S _{1/2} \rangle &=& \sum_{l_{even}} \left[ {\mathcal Y}^0_{ll}({\hat {\bf p}}
{\hat {\bf q}}) \mid (0 \frac{1}{2}) \frac{1}{2} \rangle \right] ^{\frac{1}{2} m} \:
\Psi_{ll00 \frac{1}{2}}(p q) + \sum_{l_{odd}} \left[ {\mathcal Y}^0_{ll}({\hat {\bf p}} {\hat {\bf q}}) \mid (1
\frac{1}{2}) \frac{1}{2} \rangle \right] ^{\frac{1}{2} m} \: \Psi_{ll01
\frac{1}{2}}(p q).
\label{eq:2.3}
\end{eqnarray}
The first spin state occurring in Eq.~(\ref{eq:2.3}) will be denoted as $\mid \chi^m \rangle$.
It carries the correct spin
quantum numbers of the 3N bound state and is explicitly given as
\begin{equation}
\mid \chi^m \rangle \equiv \mid (0 \frac{1}{2}) \frac{1}{2} \rangle = 
  \frac{1}{\sqrt{2}} \left( \chi^+_1 \chi^-_2 - \chi^-_1 \chi^+_2 \right) \chi^m_3 .
\label{eq:2.4}
\end{equation}
By introducing the
spin operator
\begin{equation}
\mbox{\boldmath $\sigma$} (12) \equiv \frac{1}{2} \left( \mbox{\boldmath $\sigma$}(1) 
- \mbox{\boldmath $\sigma$}(2)
\right),
\label{eq:2.5}
\end{equation}
which is odd under the exchange for particles 1 and 2, one can verify by
straightforward calculation that
\begin{equation}
\mbox{\boldmath $\sigma$} (12) \cdot \mbox{\boldmath $\sigma$}(3) 
\mid (0 \frac{1}{2})\frac{1}{2} m \rangle =
-\sqrt{3} \mid (1 \frac{1}{2}) \frac{1}{2} m \rangle.
\label{eq:2.6}
\end{equation}
This leads to the second spin state in Eq.~(\ref{eq:2.3}).
Thus, taking the antisymmetry with respect to nucleons 1 and 2 into account,
Eq.~(\ref{eq:2.3}) can be rewritten as 
\begin{eqnarray}
\mid \, ^2 S _{1/2} \rangle &=& \mid \chi^m \rangle \sum_{l_{even}} \frac{\sqrt{2l+1}}{4\pi}
  P_l({\hat {\bf p}}\cdot {\hat {\bf q}}) \: \Psi_{ll00\frac{1}{2}}(p q) \nonumber \\
& & + \frac{1}{\sqrt{3}} \mbox{\boldmath $\sigma$} (12) \cdot \mbox{\boldmath
$\sigma$}(3) \mid \chi^m
\rangle \sum_{l_{odd}} \frac{\sqrt{2l+1}}{4\pi} P_l({\hat {\bf p}}\cdot {\hat {\bf q}}) \:
\Psi_{ll01 \frac{1}{2}}(p q).
\label{eq:2.7}
\end{eqnarray}
Here the standard relation of $ {\mathcal Y}^0_{ll}({\hat {\bf p}}{\hat {\bf q}})$ to the
Legendre polynomial $P_l ({\hat {\bf p}}\cdot {\hat {\bf q}})$ has been used. These are the
first two examples for the operator form of the $^3$H bound state. The state $\mid
\chi^m \rangle$ with its correct spin quantum numbers for $^3$H is multiplied by
scalar functions and occurs either by itself or is acted upon by a rotational
invariant expression formed out of spin operators. 
Below additional rotational invariant expressions will appear in the
$^3$H wave function, which are formed out of spin operators and momentum vectors. 

The second part of Eq.~(\ref{eq:2.2}), $^2P _{1/2}$, has the following explicit form
\begin{eqnarray}
\mid \, ^2P _{1/2} \rangle &\equiv&  \sum_{l \lambda} \left[ {\mathcal Y}^1_{l
\lambda} ({\hat {\bf p}} {\hat {\bf q}}) \mid (0 \frac{1}{2}) \frac{1}{2} \rangle \right]
^{\frac{1}{2} m} \: \Psi_{l \lambda 10\frac{1}{2}}(p q) + \sum_{l \lambda} \left[ {\mathcal Y}^1_{l \lambda} ({\hat {\bf p}} {\hat {\bf q}})
\mid (1 \frac{1}{2}) \frac{1}{2} \rangle \right] ^{\frac{1}{2} m} \: \Psi_{l
\lambda 1 1 \frac{1}{2}}(p q) \nonumber \\
&\equiv& \mid ^2P _{1/2}^m \rangle \mid_{s=0} + \mid ^2P _{1/2}^m \rangle
\mid_{s=1} .
\label{eq:2.8}
\end{eqnarray}
First let us consider  $\mid ^2P _{1/2}^m \rangle \mid_{s=0}$,
insert the Clebsch-Gordon coefficients and use the standard descending spin
operator, which is related to the spherical component $\sigma_- (3)$\cite{roseangmom}. 
Doing this one obtains
\begin{equation}
\mid \, ^2P _{1/2}^{m=\frac{1}{2}} \rangle \mid_{s=0} = \frac{1}{\sqrt{3}} \sum_{l_{even \geq 2}}
 \left( -\sigma_0(3){\mathcal Y}^{10}_{ll} ({\hat {\bf p}} {\hat {\bf q}}) + 
\sigma_- (3) {\mathcal Y}^{11}_{ll} ({\hat {\bf p}} {\hat {\bf q}}) \right) \mid \chi^{m=\frac{1}{2}}
\rangle \: \Psi_{ll10\frac{1}{2}}(p q).
\label{eq:2.9}
\end{equation}
In the last equation we also used the fact that $^3$H has positive parity, which
enforces $l=\lambda$. Next we need to face the problem of the infinite sum over
$l$, which includes the angular dependence on ${\bf p}$ and ${\bf q}$. As shown
in the Appendix A, the following relation holds
\begin{equation}
{\mathcal Y}^{1m}_{ll} ({\hat {\bf p}} {\hat {\bf q}}) = c_l({\hat {\bf p}} \cdot {\hat {\bf q}}) {\mathcal
Y}^{1m}_{11} ({\hat {\bf p}} {\hat {\bf q}}),
\label{eq:2.10}
\end{equation}
where the coefficients $c_l({\hat {\bf p}} \cdot {\hat {\bf q}})$ are analytically known
functions.
Thus, we obtain for Eq.~(\ref{eq:2.9}) the expression
\begin{eqnarray}
\mid \, ^2P _{1/2}^{m=\frac{1}{2}} \rangle \mid_{s=0} & = & \frac{1}{\sqrt{3}} 
\left( -\sigma_0(3){\mathcal Y}^{10}_{11} ({\hat {\bf p}} {\hat {\bf q}}) + \sigma_- (3) {\mathcal
Y}^{11}_{11} ({\hat {\bf p}} {\hat {\bf q}}) \right) \mid \chi^{m=\frac{1}{2}} \rangle \nonumber\\
 & & \times \sum_{l_{even \geq 2}} c_l({\hat {\bf p}} \cdot {\hat {\bf q}}) \: \Psi_{ll10\frac{1}{2}}(p q). \label{eq:2.11}
\end{eqnarray}
Further one recognizes that ${\mathcal Y}^{1m}_{11} ({\hat {\bf p}} {\hat {\bf q}})$ can be
expressed in terms of the spherical components of the cross product 
${\bf p} \times {\bf q}$, which turns Eq.~(\ref{eq:2.11}) into the operator form 
\begin{equation}
\mid ^2P _{1/2}^m \rangle \mid_{s=0} = \frac{1}{4\pi} \sqrt{\frac{3}{2}}
\frac{1}{i} \frac{ \mbox{\boldmath $\sigma$}(3) \cdot {\bf p} \times {\bf q}}
{pq} \mid \chi^m \rangle \: \sum_{l_{even \geq 2}} c_l({\hat {\bf p}} \cdot
{\hat {\bf q}}) \: \Psi_{ll10\frac{1}{2}}(p q).
\label{eq:2.12}
\end{equation}

The other P-component of Eq.~(\ref{eq:2.8}), $\mid \, ^2 P^m _{1/2} \rangle \mid_{s=1}$ is a
little more complicated. Inserting the Clebsch-Gordon coefficients and making
use of the relation in Eq.~(\ref{eq:2.6}) gives
\begin{eqnarray}
\mid \, ^2P _{1/2}^{m=\frac{1}{2}} \rangle \mid_{s=1} & = & \frac{1}{3} \mbox{\boldmath $\sigma$}(12)
\cdot \mbox{\boldmath $\sigma$}(3) \nonumber\\
 & & \times \sum_{l_{odd}} \Psi_{ll11\frac{1}{2}}(p q) 
\left( \sigma_0 (3) {\mathcal Y}^{10}_{ll} ({\hat {\bf p}} {\hat {\bf q}}) - \sigma_- (3)
{\mathcal Y}^{11}_{ll} ({\hat {\bf p}} {\hat {\bf q}}) \right) 
\mid \chi^{m=\frac{1}{2}} \rangle. \label{eq:2.13}
\end{eqnarray}
Using again the relation given in  Eq.~(\ref{eq:2.10}) and expressing the
quantity ${\mathcal Y}^{1m}_{11} ({\hat {\bf p}} {\hat {\bf q}})$ in terms of the spherical
components of ${\bf p} \times {\bf q}$ leads to the intermediate result
\begin{equation}
\mid \, ^2P _{1/2}^m \rangle \mid_{s=1} = \frac{1}{4\pi} \frac{1}{\sqrt{2}} 
\sum_{l_{odd}} c_l({\hat {\bf p}} \cdot {\hat {\bf q}}) \: \Psi_{ll11\frac{1}{2}}(p q)
 i \mbox{\boldmath $\sigma$}(12) \cdot \mbox{\boldmath $\sigma$}(3)
 \frac{ \mbox{\boldmath $\sigma$}(3) \cdot {\bf p} \times {\bf q}} {pq} \mid \chi^m \rangle,
\label{eq:2.14}
\end{equation}
and finally to 
\begin{eqnarray}
\mid \, ^2P _{1/2}^m \rangle \mid_{s=1} &=& \frac{1}{4\pi} \frac{1}{\sqrt{2}}
\sum_{l_{odd}} \Psi_{ll11\frac{1}{2}}(p q) c_l({\hat {\bf p}} \cdot {\hat {\bf q}}) \nonumber \\
 & & \times \frac{1} {pq}  \left(  i\mbox{\boldmath $\sigma$}(12) \cdot {\bf p}
\times {\bf q} - (\mbox{\boldmath $\sigma$}(3) \times \mbox{\boldmath
$\sigma$}(12) ) \cdot ({\bf p} \times {\bf q}) \right) \mid \chi^m \rangle.
\label{eq:2.15}
\end{eqnarray}

The next term to calculate from Eq.~(\ref{eq:2.2}) is $\mid \, ^4P _{1/2}^m
\rangle$. Again we insert the Clebsch-Gordon coefficients and find
\begin{eqnarray}
\mid \, ^4P _{1/2}^{m=1/2} \rangle &=& \frac{1}{\sqrt{2}} \mid (1 \frac{1}{2})
\frac{3}{2} \frac{3}{2} \rangle \sum_{l_{odd}} {\mathcal Y}^{1 -1}_{ll} ({\hat
{\bf p}} {\hat {\bf q}}) \: \Psi_{ll 1 1 \frac{3}{2}} (p q) \nonumber \\
& & - \frac{1}{\sqrt{3}} \mid (1 \frac{1}{2}) \frac{3}{2} \frac{1}{2} \rangle
\sum_{l_{odd}} {\mathcal Y}^{1 0}_{ll} ({\hat {\bf p}} {\hat {\bf q}}) \: \Psi_{ll 1 1
\frac{3}{2}} (p q) \nonumber \\ 
& & + \frac{1}{\sqrt{6}} \mid (1 \frac{1}{2}) \frac{3}{2} \frac{-1}{2} \rangle
\sum_{l_{odd}} {\mathcal Y}^{1 1}_{ll} ({\hat {\bf p}} {\hat {\bf q}}) \: \Psi_{ll 1 1
\frac{3}{2}} (p q).
\label{eq:2.16}
\end{eqnarray}
For the sake of a simpler notation we used $m=1/2$. As shown in Appendix B,
this can be cast into the form
\begin{eqnarray}
\mid \, ^4P _{1/2}^{m=1/2} \rangle &=& \frac{1}{i} \sum_{l_{odd}} c_l({\hat {\bf p}} \cdot
{\hat {\bf q}}) \: \Psi_{ll 1 1 \frac{3}{2}} (p q) \nonumber\\
 & & \times \frac{1}{4\pi} \left( \left[ \mbox{\boldmath $\sigma$}(12) - \frac{i}{2}
(\mbox{\boldmath $\sigma$}(3) \times \mbox{\boldmath$\sigma$}(12) ) \right] 
\cdot \frac{ {\bf p} \times {\bf q} }{pq} \right) \mid \chi^{m=1/2} \rangle. \label{eq:2.17} 
\end{eqnarray}
Of course, this relation is also valid for $m=-1/2$.
 
Finally, we turn to the last part of Eq.~(\ref{eq:2.2}), the expression for
$^4D _{1/2}^{m=1/2}$. Again inserting the Clebsch-Gordon coefficients yields
\begin{eqnarray}
\mid \, ^4D _{1/2}^{m=1/2} \rangle &=& -\frac{1}{\sqrt{10}} \mid (1 \frac{1}{2})
\frac{3}{2} \frac{3}{2} \rangle \sum_{l \lambda} {\mathcal Y}^{2 -1}_{l \lambda}
({\hat {\bf p}} {\hat {\bf q}}) \: \Psi_{l \lambda 21 \frac{3}{2}} (p q) \nonumber \\
& & + \frac{1}{\sqrt{5}} \mid (1 \frac{1}{2}) \frac{3}{2} \frac{1}{2} \rangle
\sum_{l \lambda} {\mathcal Y}^{2 0}_{l \lambda} ({\hat {\bf p}} {\hat {\bf q}}) \: \Psi_{l
\lambda 21 \frac{3}{2}} (p q) \nonumber \\
& & - \sqrt{\frac{3}{10}} \mid (1 \frac{1}{2}) \frac{3}{2} -\frac{1}{2} \rangle
\sum_{l \lambda} {\mathcal Y}^{2 1}_{l \lambda} ({\hat {\bf p}} {\hat {\bf q}}) \: \Psi_{l
\lambda 21 \frac{3}{2}} (p q) \nonumber \\
& & + \sqrt{\frac{2}{5}} \mid (1 \frac{1}{2}) \frac{3}{2} -\frac{3}{2} \rangle
\sum_{l \lambda} {\mathcal Y}^{2 2}_{l \lambda} ({\hat {\bf p}} {\hat {\bf q}}) \: \Psi_{l
\lambda 21 \frac{3}{2}} (p q).
\label{eq:2.18}
\end{eqnarray}
Due to the overall positive parity and the antisymmetry of the state with respect to the two
neutrons the sum over $l$ and $\lambda$ splits as
\begin{equation}
\sum_{l \lambda} {\mathcal Y}^{2 \mu}_{l \lambda} \: \Psi_{l \lambda 21 \frac{3}{2}}
= \sum_{l_{odd}} {\mathcal Y}^{2 \mu}_{ll} \: \Psi_{ll21\frac{3}{2}}
  + \sum_{l_{odd}} {\mathcal Y}^{2 \mu}_{l \, l+2} \: \Psi_{l \, l+2 \,
21\frac{3}{2}} + \sum_{l_{odd}} {\mathcal Y}^{2 \mu}_{l+2 \, l} \: \Psi_{l+2 \, l \,
21\frac{3}{2}}.
\label{eq:2.19}
\end{equation}
Now, however, the different types of coupled spherical harmonics are more difficult
to split into simple second order expressions and scalar functions. This is
elaborated in the Appendix C with the result
\begin{eqnarray}
{\mathcal Y}^{2 \mu}_{ll} ({\hat {\bf p}} {\hat {\bf q}}) &=& {\mathcal Y}^{2 \mu}_{11} ({\hat
{\bf p}} {\hat {\bf q}}) \: A_l + B_l \left( Y_{2\mu} ({\hat {\bf p}}) + Y_{2\mu} ({\hat {\bf q}}) \right) 
  \nonumber \\
{\mathcal Y}^{2 \mu}_{l \, l+2}({\hat {\bf p}} {\hat {\bf q}}) &=&{\mathcal Y}^{2 \mu}_{11}
({\hat {\bf p}} {\hat {\bf q}}) C_l +D_l \: Y_{2\mu} ({\hat {\bf q}}) +E_l Y_{2\mu} ({\hat {\bf p}})
\nonumber \\
{\mathcal Y}^{2 \mu}_{l+2 \, l} ({\hat {\bf p}} {\hat {\bf q}})  &=& {\mathcal Y}^{2 \mu}_{11}
({\hat {\bf p}} {\hat {\bf q}})  C_l + D_l Y_{2\mu} ({\hat {\bf p}}) +E_l Y_{2\mu} ({\hat {\bf q}}).
\label{eq:2.20}
\end{eqnarray}
The quantities $A_l, \: \ldots , \: E_l$ are analytically known scalar functions depending
on the momenta ${\bf p}$ and ${\bf q}$. They can be inferred from Appendix C and the first relevant ones are given below in Eqs.(\ref{eq:2.28}) 
and (\ref{eq:2.29}). Using the decompositions of
Eq.~(\ref{eq:2.20}) in Eq.~(\ref{eq:2.19}) leads to
\begin{equation}
\sum_{l \lambda} {\mathcal Y}^{2 \mu}_{l \lambda} \: \Psi_{l \lambda 21 \frac{3}{2}}
= {\mathcal Y}^{2 \mu}_{11} ({\hat {\bf p}} {\hat {\bf q}})  \sum_{l_{odd}} X_l +
Y_{2\mu}({\hat {\bf p}}) \sum_{l_{odd}} V_l + Y_{2\mu}({\hat {\bf q}}) \sum_{l_{odd}} W_l ,
\label{eq:2.23}
\end{equation}
with 
\begin{eqnarray}
X_l &=& A_l \Psi_{ll \, 21 \frac{3}{2}} + C_l \left( \Psi_{l \, l+2\, 21
\frac{3}{2}} + \Psi_{l+2 \, l \, 21 \frac{3}{2}} \right)  \\
V_l &=& B_l \Psi_{ll \, 21 \frac{3}{2}}  +E_l \Psi_{l \, l+2\, 21\frac{3}{2}} +D_l
\Psi_{l+2 \, l \, 21 \frac{3}{2}} \\
W_l &=& B_l \Psi_{ll \, 21 \frac{3}{2}} + D_l \Psi_{l \, l+2\, 21\frac{3}{2}} +E_l
\Psi_{l+2 \, l \, 21 \frac{3}{2}} .
\label{eq:2.26}
\end{eqnarray}
 
For the representation of the spin states $\mid (1 \frac{1}{2})\frac{3}{2}
M_S \rangle$ of Eq.~(\ref{eq:2.18}) in terms of the spin state $\mid \chi^m \rangle$ 
we use an equivalent but modified form as shown above. The details are shown in 
Appendix D. After some algebra we arrive at
\begin{eqnarray}
\mid \, ^4D _{1/2}^{m=1/2} \rangle &=&
\frac{1}{2} \sqrt{\frac{3}{2\pi}} \left[ \frac{\mbox{\boldmath$\sigma$}(12) \cdot
{\bf p} \: \mbox{\boldmath $\sigma$}(3) \cdot {\bf p}}{p^2} - \frac{1}{3}
\mbox{\boldmath$\sigma$}(12) \cdot \mbox{\boldmath $\sigma$}(3) \right] \mid \chi^m
\rangle \: \sum_{l_{odd}} V_l \nonumber \\
& & + \frac{1}{2} \sqrt{\frac{3}{2\pi}} \left[ \frac{\mbox{\boldmath$\sigma$}(12)
\cdot {\bf q} \: \mbox{\boldmath $\sigma$}(3) \cdot {\bf q}} {q^2}  -
\frac{1}{3} \mbox{\boldmath$\sigma$}(12) \cdot \mbox{\boldmath $\sigma$}(3) \right]
\mid \chi^m \rangle \: \sum_{l_{odd}} W_l \nonumber \\
& & + \frac{1}{2} \frac{3}{4\pi}  \frac{1}{\sqrt{5}} \frac{1}{pq} 
\Bigl[ \mbox{\boldmath$\sigma$}(12) \cdot {\bf q} \: \mbox{\boldmath $\sigma$}(3) \cdot {\bf p} 
+ \mbox{\boldmath$\sigma$}(12) \cdot {\bf p} 
\: \mbox{\boldmath $\sigma$}(3) \cdot {\bf q} \nonumber\\
& & \qquad \qquad \qquad - \frac{2}{3} {\bf p} \cdot {\bf q}
\mbox{\boldmath$\sigma$}(12) \cdot \mbox{\boldmath $\sigma$}(3)\Bigr] \mid
\chi^m\rangle \: \sum_{l_{odd}} X_l . \label{eq:2.27} 
\end{eqnarray}

It is now the time to compare our results to the scalar expressions given in
\cite{gerjuoy42}. We see that the first eight terms of Eqs.~(2)-(7) in \cite{gerjuoy42}
are identical to the ones derived here (see Eqs.~(\ref{eq:2.7}),(\ref{eq:2.12}),
(\ref{eq:2.15}), (\ref{eq:2.17}), and (\ref{eq:2.27})). While in \cite{gerjuoy42} the
scalar functions multiplying the scalar  operators are unknown, here they are explicitly
provided in terms of the partial wave function components calculated e.g. in a Faddeev
approach. We want to point out that the last expression in Eq.~(7) of Ref. 
\cite{gerjuoy42} is redundant. By itself it is not antisymmetric under the exchange of
particles $1$ and $2$ (the two neutrons). It has to be multiplied by a scalar
function which is formed from odd orbital angular momenta $l$. Doing this, one
arrives after some algebra at the following result: The $L=2$ piece is already
contained in the previous three terms of $\mid \, ^4D _{1/2}^m\rangle$. The $L=1$
piece is identically zero, and the $L=0$ part cancels among the two terms given in
the last expression in Eq.~(7) of \cite{gerjuoy42}. 

\subsection{Normalization}

In the previous subsection we started from a partial wave decomposition 
of the 3N wave function $\Psi^m({\bf p}, {\bf q})$ (Eq.~(\ref{eq:2.2})). By the very
construction the individual terms are manifestly orthogonal. In their new forms, as
derived in the previous subsection, this is no longer obvious. However, since the
new forms are identical reformulations of the original terms, it is simplest to go
back to Eq.~(\ref{eq:2.2}) to verify orthogonality and normalization. Then one sees
immediately that $^2 S_{\frac{1}{2}}$, $^2 P_{\frac{1}{2}}$, and $^4
P_{\frac{1}{2}}$ given in Eqs.~(\ref{eq:2.7}), (\ref{eq:2.12}), (\ref{eq:2.15}) and (\ref{eq:2.17}) 
are orthogonal to each other. The normalization for those three
pieces is given as
\begin{eqnarray}
\langle \, ^2 S _{1/2} + \, ^2 P _{1/2} +\,  ^4 P _{1/2} \mid \, ^2 S _{1/2} +\, 
 ^2 P _{1/2} + \, ^4 P _{1/2} \rangle =  
\sum_{sSl\lambda} \sum_{L=0,1} \int dp \,  p^2 \int dq \,  q^2
\Psi^2_{l\lambda LsS}(pq).
\label{eq:2.28}
\end{eqnarray}
The last term in Eq.~(\ref{eq:2.2}), $\mid \, ^4 D _{1/2} \rangle$, is more
intricate in the form of Eq.~(\ref{eq:2.27}). 
Of course one can also go back to Eq.~(\ref{eq:2.2}).
However, instead of doing this, we go back half way to Eq.~(\ref{eq:2.18}) and insert the
decomposition given in Eq.~(\ref{eq:2.23}).
This leads to the three terms
\begin{eqnarray}
\mid \, ^4 D^m _{1/2} \rangle &=& \left[ Y_2({\hat {\bf p}}) \mid (1\frac{1}{2}) \frac{3}{2}
\rangle \right]^{\frac{1}{2} m} \sum_{l_{odd}} V_l \nonumber\\
& & + \left[ Y_2({\hat {\bf q}}) \mid (1\frac{1}{2}) \frac{3}{2} \rangle
\right]^{\frac{1}{2} m} \sum_{l_{odd}} W_l \nonumber\\
 & & +  \left[ {\mathcal Y}^2_{11}({\hat {\bf p}} {\hat {\bf q}}) 
\mid (1\frac{1}{2}) \frac{3}{2} \rangle \right]^{\frac{1}{2} m} \sum_{l_{odd}} X_l,
\label{eq:2.29}
\end{eqnarray}
which are in unique correspondence to the three terms in  Eq.~(\ref{eq:2.27}). The question of normalization and orthogonality requires knowledge of the analytically known coefficients $A_l, \: \ldots , \: E_l$ inside $V_l$, $W_l$ and $X_l$.  

As shown in Appendix C, they are given as follows.
If we keep only $l=\lambda = 1$, then 
\begin{eqnarray}
A_1 &=& 1 \nonumber \\
B_1 &=& 0 \label{eq:2.30a}
\end{eqnarray}
and the coefficients C, D, and E do not occur. If we allow in addition $l=3$ and $\lambda =3$, but no higher values, then 
\begin{eqnarray}
A_3 &=& \frac{1}{9} \sqrt{\frac{7}{2}} \left( 25 P_2 ({\hat {\bf p}} \cdot {\hat {\bf q}})+11
\right) \nonumber \\
B_3 &=& -\sqrt{\frac{35}{12\pi}} P_1 ({\hat {\bf p}} \cdot {\hat {\bf q}}) \nonumber \\
C_1 &=& \sqrt{\frac{2}{3}} \nonumber \\
D_1 &=& -\sqrt{\frac{5}{4\pi}} P_1 ({\hat {\bf p}} \cdot {\hat {\bf q}}) \nonumber\\
E_1 &=& 0 \label{eq:2.30}
\end{eqnarray}
Note that the coefficients $C_1$, $D_1$, and $E_1$ occur together with $l=3$.
The evaluation of the terms $l=5$ and higher can be found from the general formula
given in Appendix C.

If one keeps only the $l=1$ partial wave function component, then 
$\mid \, ^4 D^m _{1/2} \rangle$ reduces to the simple expression
\begin{equation}
\mid \, ^4 D^m _{1/2} \rangle \mid_{l=1} = \left[ {\mathcal Y}^2_{11}({\hat {\bf p}} {\hat
{\bf q}}) \mid (1\frac{1}{2}) \frac{3}{2} \rangle \right]^{\frac{1}{2} m}
\Psi_{1121\frac{3}{2}} (pq),
\label{eq:2.31}
\end{equation}
which is orthogonal to the previous states and normalized as
\begin{equation}
\langle \, ^4 D^m _{1/2}  \mid \, ^4 D^m _{1/2} \rangle \Big|_{l=1} = \int dp \, p^2 \int
dq \, q^2 \, \Psi^2 _{1121\frac{3}{2}} (pq).
\label{eq:2.32}
\end{equation}

If one includes the $l=3$ or $\lambda =3$ partial wave function components, which are in fact tiny
contributions, one arrives at
\begin{eqnarray}
\mid \, ^4 D _{1/2} \rangle \Big|_{l=1,3} &=& \left[ Y_2 ({\hat {\bf p}}) \mid (1\frac{1}{2})
\frac{3}{2} \rangle \right]^{\frac{1}{2} m} \left( -\frac{1}{\sqrt{4\pi}} \sqrt{5} \right)
 \left( \sqrt{\frac{7}{3}} \Psi_{3321\frac{3}{2}}(pq) + \Psi_{1321\frac{3}{2}}(pq)
\right) P_1({\hat {\bf p}} \cdot{\hat {\bf q}}) \nonumber \\
& & + \left[ Y_2 ({\hat {\bf q}}) \mid (1\frac{1}{2})\frac{3}{2} \rangle
\right]^{\frac{1}{2} m}  \left( -\frac{1}{\sqrt{4\pi}} \sqrt{5} \right) 
 \left( \sqrt{\frac{7}{3}} \Psi_{3321\frac{3}{2}}(pq) + \Psi_{3121\frac{3}{2}}(pq)
\right) P_1({\hat {\bf p}} \cdot{\hat {\bf q}}) \nonumber \\
& & + \left[ {\mathcal Y}^2_{11}({\hat {\bf p}}{\hat {\bf q}}) \mid (1\frac{1}{2})\frac{3}{2}
\rangle \right]^{\frac{1}{2} m} 
  \Bigg( \Psi_{1121\frac{3}{2}}(pq)
+\frac{11}{9}\sqrt{\frac{7}{2}} \Psi_{3321\frac{3}{2}}(pq) \nonumber \\
& & +\sqrt{\frac{2}{3}} 
\left( \Psi_{1321\frac{1}{3}}(pq) + \Psi_{3121\frac{3}{2}}(pq) \right) 
+ \frac{25}{9} \sqrt{\frac{7}{2}} \Psi_{33121\frac{1}{3}}(pq) P_2 ({\hat {\bf p}}
\cdot{\hat {\bf q}}) \Bigg).
\label{eq:2.33} 
\end{eqnarray}
Now the three terms are no longer orthogonal to each other, but of course an
identical reformulation of 
\begin{eqnarray}
\mid \, ^4 D^m _{1/2} \rangle \Big|_{l=1,3} &=& \left[{\mathcal Y}^2_{11}({\hat
{\bf p}}{\hat {\bf q}}) \mid (1\frac{1}{2})\frac{3}{2}\rangle \right]^{\frac{1}{2}
m}\Psi_{1121\frac{3}{2}}(pq) \nonumber \\ 
& & + \left[ {\mathcal Y}^2_{33}({\hat {\bf p}}{\hat {\bf q}}) \mid
(1\frac{1}{2})\frac{3}{2}\rangle \right]^{\frac{1}{2}m} \Psi_{3321\frac{3}{2}}(pq) 
+\left[ {\mathcal Y}^2_{13}({\hat {\bf p}}{\hat {\bf q}}) \mid (1\frac{1}{2})\frac{3}{2}\rangle
\right]^{\frac{1}{2}m} \Psi_{1321\frac{3}{2}}(pq) \nonumber \\
& & + \left[ {\mathcal Y}^2_{31}({\hat {\bf p}}{\hat {\bf q}}) \mid
(1\frac{1}{2})\frac{3}{2}\rangle \right]^{\frac{1}{2}m} \Psi_{3121\frac{3}{2}}(pq).
\label{eq:2.34}
\end{eqnarray}
as given in Eq.~(\ref{eq:2.2}).
Consequently, the state $\mid \, ^4 D^m _{1/2} \rangle$ is normalized as
\begin{eqnarray}
\langle \, ^4 D^m _{1/2} \mid \, ^4 D^m _{1/2} \rangle \Big|_{l=1,3} &=& 
\int dp \, p^2 \int dq \, q^2 \Bigg( \Psi^2_{1121\frac{3}{2}}(pq) 
+ \Psi^2_{3321\frac{3}{2}}(pq) \nonumber\\
& & \qquad \qquad \qquad \quad + \Psi^2_{1321\frac{3}{2}}(pq) 
+ \Psi^2_{3121\frac{3}{2}}(pq) \Bigg) . \label{eq:2.35}
\end{eqnarray}
The direct verification of this result in form of Eq.~(\ref{eq:2.27}) is straightforward but
tedious.

Summarizing this section, the sum of the expressions in Eqs.~(\ref{eq:2.7}), (\ref{eq:2.12}), 
(\ref{eq:2.15}), (\ref{eq:2.17}), and (\ref{eq:2.27}) is the operator form of the 3N bound state in
momentum space we were looking for. 
It has the form 
\begin{equation}
\Psi^m({\bf p}, {\bf q}) = \sum_{i=1}^8 \phi_i ({\bf p}, {\bf q}) |\chi_i\rangle ,
\label{eq:3.1}
\end{equation}
where the $|\chi_i\rangle$ are composed out of 8 scalar operators acting on the special 3N spin state $|\chi^m \rangle$, introduced in Eq.~(\ref{eq:2.4}). For the convenience of the reader, we list the states $|\chi_i\rangle$ again. 
\begin{eqnarray}
|\chi_1\rangle &=& |\chi^m\rangle \nonumber\\
|\chi_2\rangle &=& \frac{1}{\sqrt{3}} \mbox{\boldmath $\sigma$} (12) \cdot \mbox{\boldmath
  $\sigma$}(3) \mid \chi^m \rangle \nonumber\\
|\chi_3\rangle &=& \sqrt{\frac{3}{2}}
\frac{1}{i} \frac{ \mbox{\boldmath $\sigma$}(3) \cdot {\bf p} \times {\bf q}}
{pq} \mid \chi^m \rangle \nonumber\\
|\chi_4\rangle &=& \frac{1}{\sqrt{2}}
 \frac{1} {pq}  \left(  i\mbox{\boldmath $\sigma$}(12) \cdot {\bf p}
\times {\bf q} - (\mbox{\boldmath $\sigma$}(3) \times \mbox{\boldmath
$\sigma$}(12) ) \cdot ({\bf p} \times {\bf q}) \right) \mid \chi^m \rangle \nonumber\\
|\chi_5\rangle &=& \frac{1}{i} \left( \left[ 
\mbox{\boldmath $\sigma$}(12) - \frac{i}{2}
(\mbox{\boldmath $\sigma$}(3) \times \mbox{\boldmath$\sigma$}(12) ) \right]
\cdot \frac{ {\bf p} \times {\bf q} }{pq} \right) \mid
\chi^m \rangle  \nonumber\\
|\chi_6\rangle &=& \sqrt{\frac{3}{2}} \left[ \frac{\mbox{\boldmath$\sigma$}(12) \cdot
{\bf p} \: \mbox{\boldmath $\sigma$}(3) \cdot {\bf p}}{p^2} - \frac{1}{3}
\mbox{\boldmath$\sigma$}(12) \cdot \mbox{\boldmath $\sigma$}(3) \right] \mid \chi^m
\rangle \nonumber\\
|\chi_7\rangle &=& \sqrt{\frac{3}{2}} \left[ \frac{\mbox{\boldmath$\sigma$}(12)
\cdot {\bf q} \: \mbox{\boldmath $\sigma$}(3) \cdot {\bf q}} {q^2}  -
\frac{1}{3} \mbox{\boldmath$\sigma$}(12) \cdot \mbox{\boldmath $\sigma$}(3) \right]
\mid \chi^m \rangle \nonumber\\
|\chi_8\rangle &=& \frac{3}{2}  \frac{1}{\sqrt{5}} \frac{1}{pq} \left[ \mbox{\boldmath$\sigma$}(12) \cdot {\bf q} \: \mbox{\boldmath $\sigma$}(3)
\cdot {\bf p} + \mbox{\boldmath$\sigma$}(12) \cdot {\bf p} \: \mbox{\boldmath
$\sigma$}(3) \cdot {\bf q} - \frac{2}{3} {\bf p} \cdot {\bf q}
\mbox{\boldmath$\sigma$}(12) \cdot \mbox{\boldmath $\sigma$}(3)\right] \mid
\chi^m\rangle \label{eq:2.36}
\end{eqnarray}
Each term is composed of scalar operators consisting of spin
operators and momentum vectors, applied on the pure spin state $\mid \chi^m \rangle$, which 
carries the overall quantum number $J = 1/2$. 
Furthermore, each of those terms  in Eq.~(\ref{eq:3.1}) includes scalar functions $\phi_i$ formed
out of the two Jacobi momenta ${\bf p}$ and ${\bf q}$. Their dependence on the standard partial
wave function components has been explicitly worked out and will be investigated in the following
Section III. Of course, exactly the same
forms are valid
in configuration space. In this case the Jacobi momenta ${\bf p}$ and ${\bf q}$ would have to be
replaced by the corresponding conjugate configuration space Jacobi vectors.

\section{The Scalar Functions}
The operator form of the 3N bound state  as given in Eq.~(\ref{eq:3.1}) 
contains the scalar functions $ \phi_i$. They will be investigated now and
are given according to Eqs.~(\ref{eq:2.7}), (\ref{eq:2.12}), (\ref{eq:2.15}), (\ref{eq:2.17}), and
(\ref{eq:2.27}) by
\begin{eqnarray}
\phi_1 ({\bf p}, {\bf q}) &=& \frac{1}{4\pi} \sum_{l_{even}} \sqrt{2l+1} \: P_l ({\hat {\bf p}} {\hat {\bf q}})
\:  \Psi_{ll00\frac{1}{2}}(pq) \nonumber \\
\phi_2 ({\bf p}, {\bf q}) &=& \frac{1}{4\pi} \sum_{l_{odd}} \sqrt{2l+1} \: P_l ({\hat {\bf p}} {\hat {\bf q}})
\: \Psi_{ll01\frac{1}{2}}(pq) \nonumber \\
\phi_3 ({\bf p}, {\bf q}) &=& \frac{1}{4\pi} \sum_{l_{even, l\geq 2}} c_l ({\hat {\bf p}} \cdot {\hat {\bf q}}) 
\: \Psi_{ll10\frac{1}{2}}(pq) \nonumber \\
\phi_4 ({\bf p}, {\bf q}) &=& \frac{1}{4\pi} \sum_{l_{odd}} c_l ({\hat {\bf p}} \cdot {\hat {\bf q}})
\: \Psi_{ll11\frac{1}{2}}(pq) \nonumber \\
\phi_5 ({\bf p}, {\bf q}) &=& \frac{1}{4\pi} \sum_{l_{odd}} c_l ({\hat {\bf p}} \cdot {\hat {\bf q}})
\: \Psi_{ll11\frac{3}{2}}(pq) \nonumber \\
\phi_6 ({\bf p}, {\bf q}) &=& \frac{1}{2\sqrt{\pi}} \sum_{l_{odd}} 
\left( B_l \: \Psi_{ll21\frac{3}{2}}(pq) +E_l
\: \Psi_{l\, l+2\, 21\frac{3}{2}}(pq) +D_l \: \Psi_{l+2\, l\, 21\frac{3}{2}}(pq) \right)
\nonumber \\
\phi_7 ({\bf p}, {\bf q}) &=& \frac{1}{2\sqrt{\pi}} \sum_{l_{odd}} 
\left( B_l\: \Psi_{ll21\frac{3}{2}}(pq) +D_l
\: \Psi_{l\, l+2\, 21\frac{3}{2}}(pq) +E_l \: \Psi_{l+2\, l\, 21\frac{3}{2}}(pq) \right) \nonumber \\
\phi_8 ({\bf p}, {\bf q}) &=&\frac{1}{4\pi} \sum_{l_{odd}} \left( A_l \: \Psi_{ll21\frac{3}{2}}(pq)
+C_l \left( \Psi_{l\, l+2\, 21\frac{3}{2}}(pq) + \Psi_{l+2\, l\, 21\frac{3}{2}}(pq) \right) \right)
\label{eq:3.2}
\end{eqnarray}

\begin{figure}
\centerline{\epsfig{file=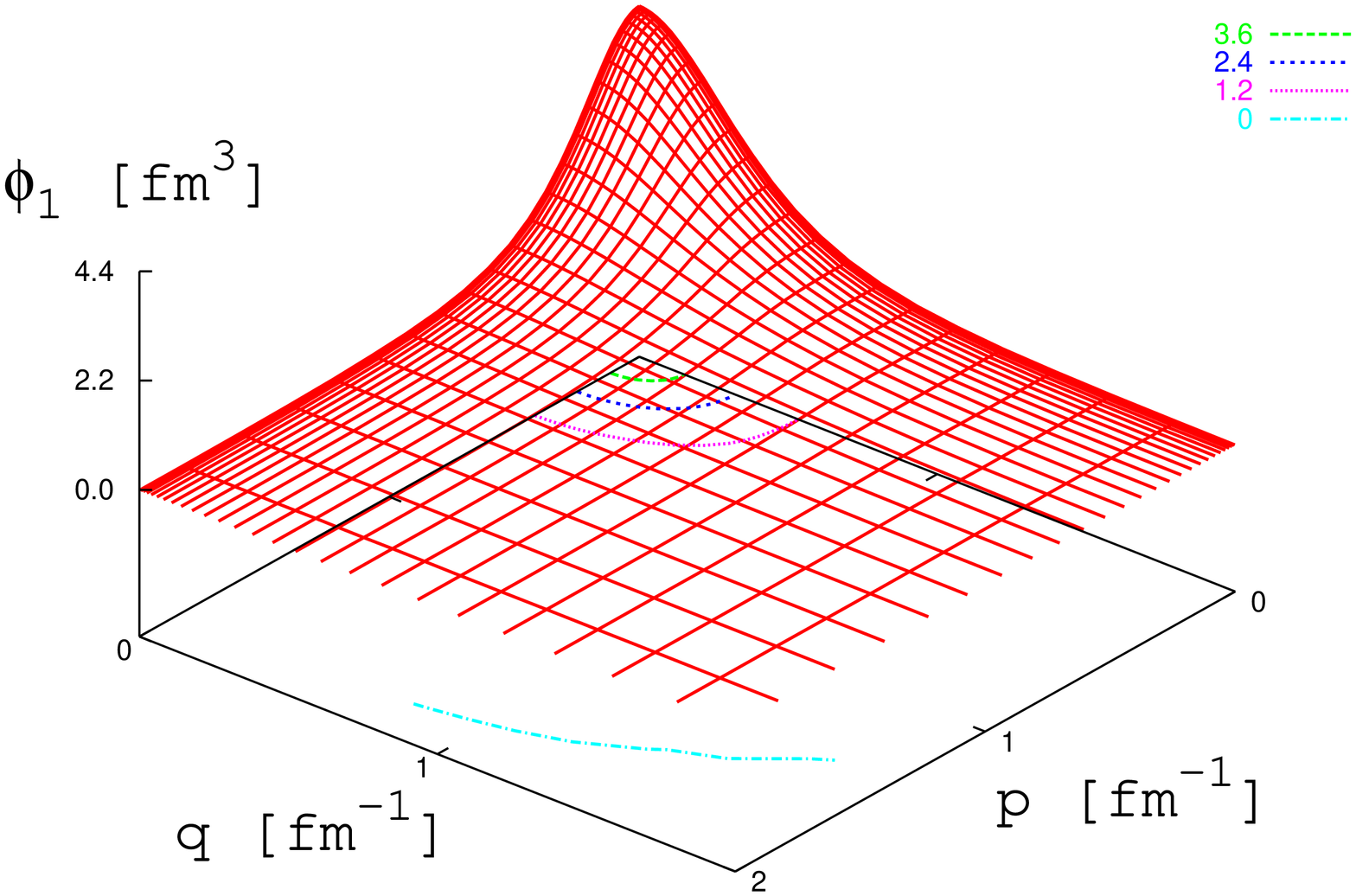,width=90mm,angle=-90}}
\caption{ The dependence of the function $\phi_1 (p,q,x=1)$ on $p$ and $q$. For better visibility also contour lines are given. 
\label{fig1}}
\end{figure}

\begin{figure}
\centerline{\epsfig{file=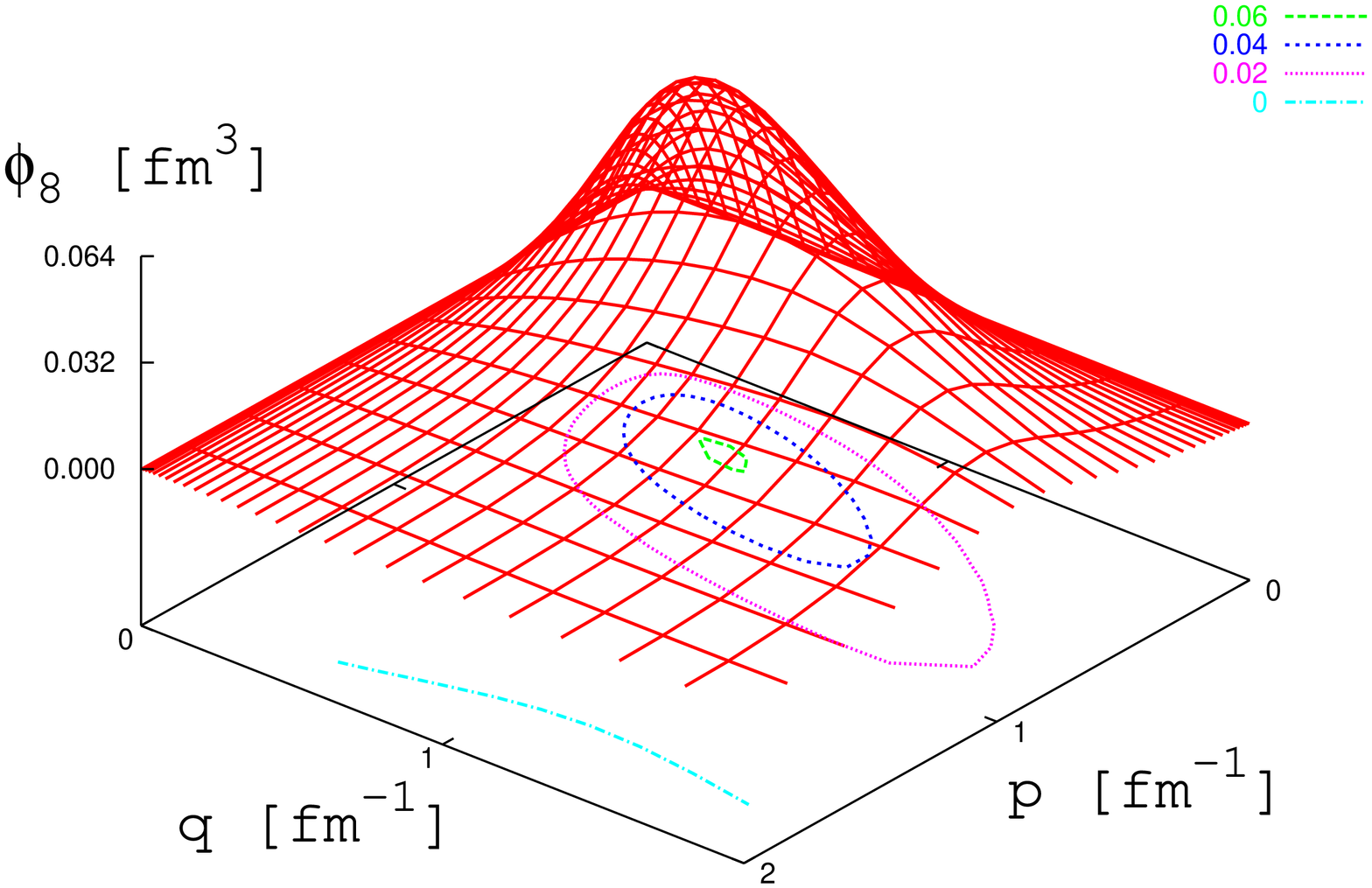,width=90mm,angle=-90}}
\caption{ The dependence of the function $\phi_8  (p,q,x=1)$ on $p$ and $q$. For better visibility also contour lines are given. 
\label{fig2}}
\end{figure}

\begin{figure}
\centerline{\epsfig{file=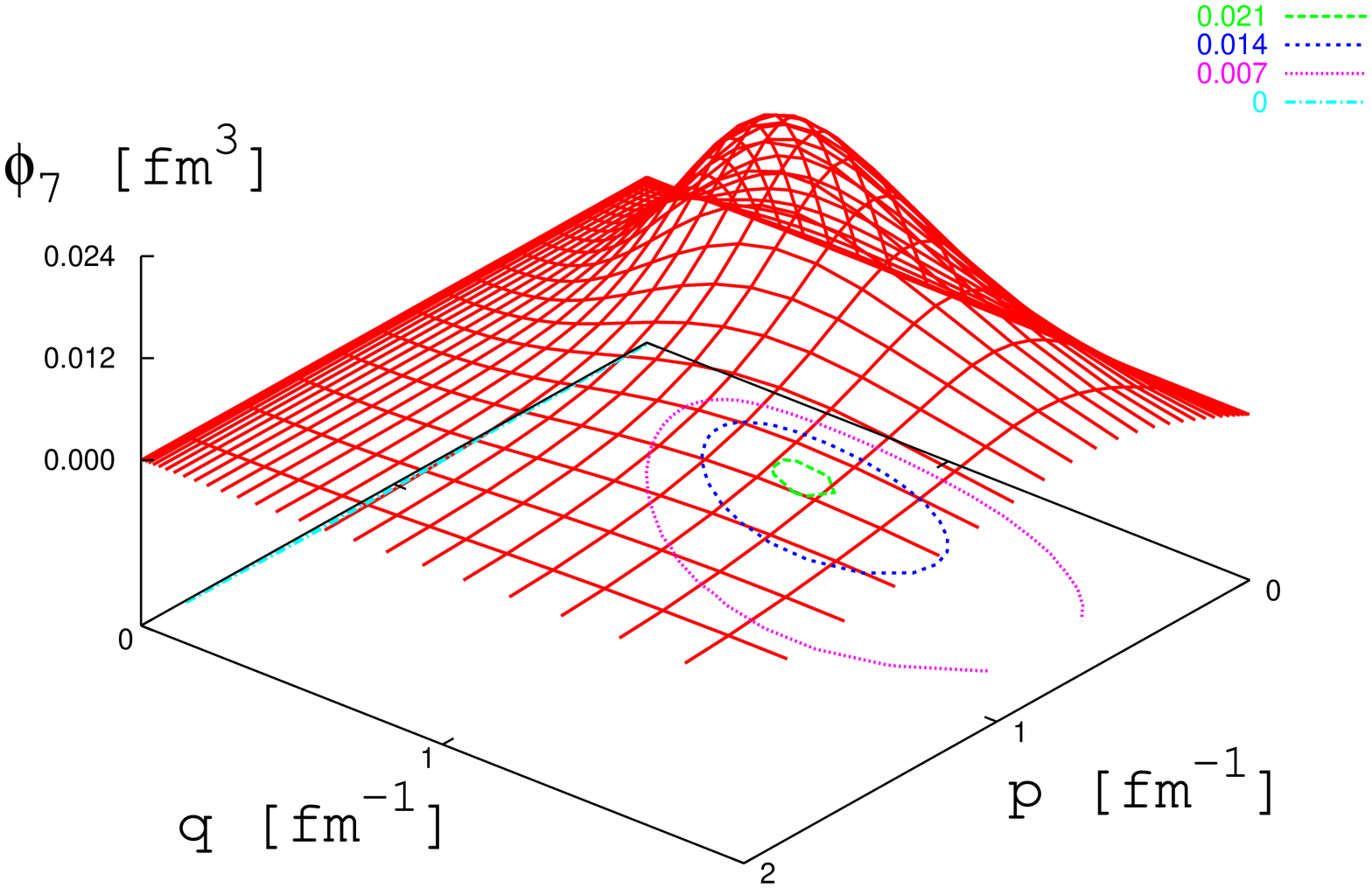,width=90mm,angle=-90}}
\caption{ The dependence of the function $\phi_7 (p,q,x=1)$ on $p$ and $q$. For better visibility also contour lines are given. 
\label{fig3}}
\end{figure}

\begin{figure}
\centerline{\epsfig{file=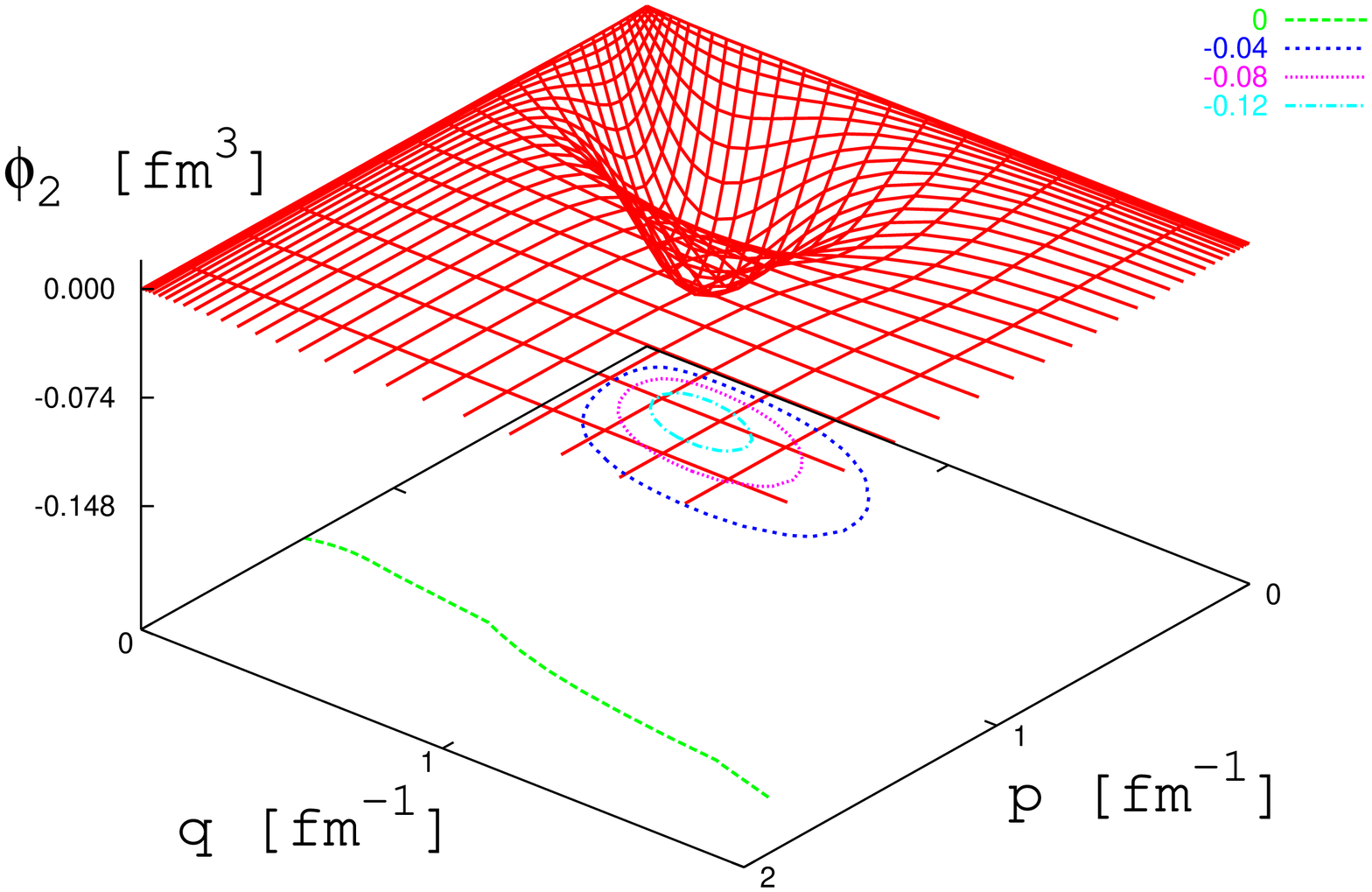,width=90mm,angle=-90}}
\caption{ The dependence of the function $\phi_2 (p,q,x=1)$ on $p$ and $q$. For better visibility also contour lines are given. 
\label{fig4}}
\end{figure}

These functions $\phi_i ({\bf p}, {\bf q})$ depend on three variables, $p$, $q$, and ${\bf {\hat p}}\cdot {\bf {\hat q}}$. They are determined by the partial wave components of the 3N
bound state and analytically known coefficient functions. In Figs.~\ref{fig1} through 
\ref{fig4} we display $\phi_1
(p,q,\cos \theta)$, $\phi_8 (p,q,\cos \theta)$, $\phi_7 (p,q,\cos \theta)$,and  $\phi_2 (p,q,\cos
\theta)$ for a fixed angle $\theta = 0$. We see that the numerically largest function is $\phi_1
(p,q,\cos \theta)$, the other three ones shown are at least an order of magnitude smaller.  
While $\phi_1 (p,q,\cos \theta)$ has a simple, bell-like shape with the maximum at
$p=0$ and $q=0$, for $\phi_8 (p,q,\cos \theta)$ the maximum is shifted to $p \approx 0.2 $ and
$q \approx 0.4 $ fm$^{-1}$. 
The reason for this is that $\phi_8 (p,q,\cos \theta)$ does not contain 
s-wave contributions but instead includes tensor force couplings. The function
$\phi_7 (p,q,\cos \theta)$ is similar in shape to $\phi_8 (p,q,\cos \theta)$. Finally, 
$\phi_2 (p,q,\cos \theta)$ also has its minimum shifted away from the origin. 

\begin{figure}
\centerline{\epsfig{file=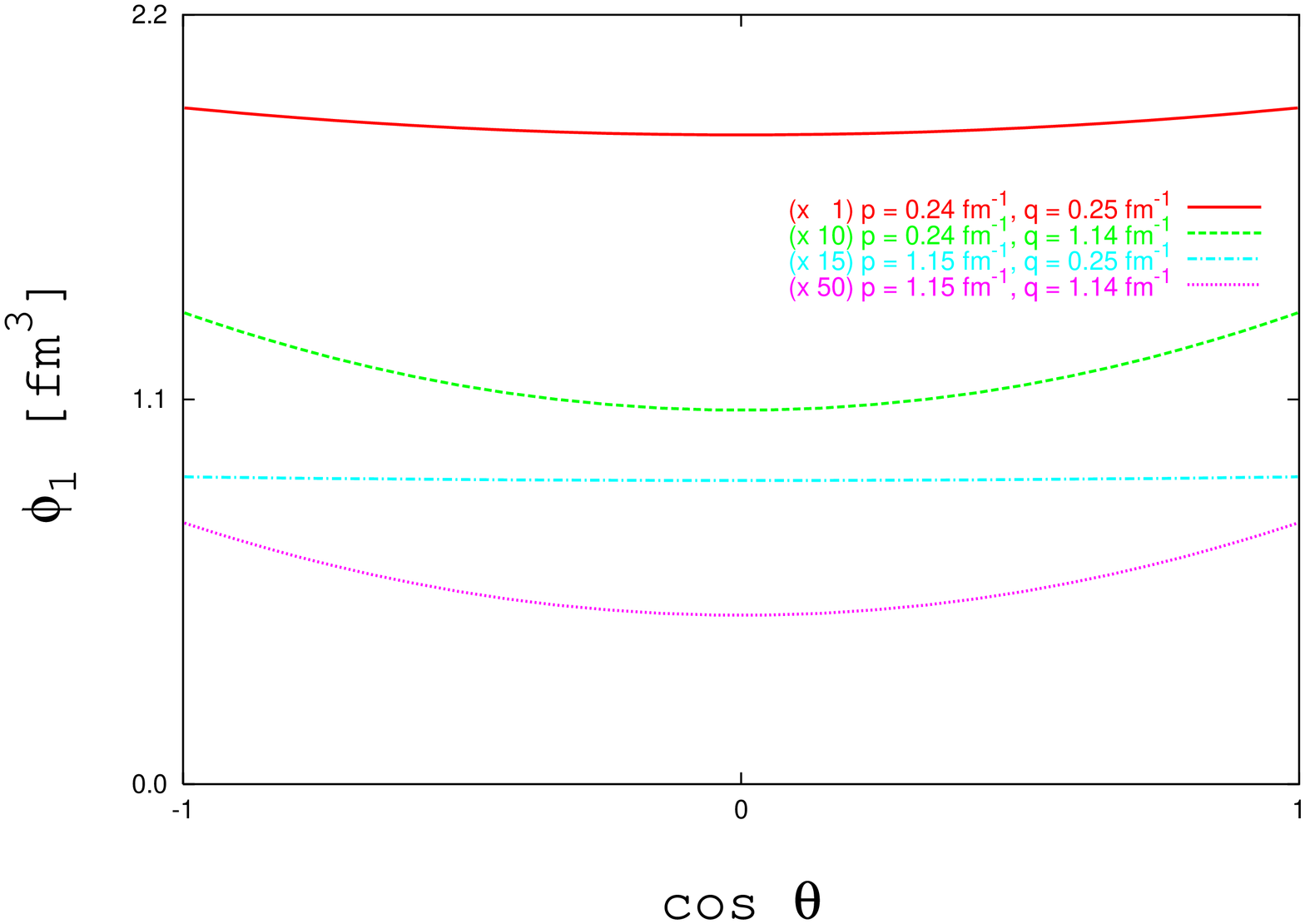,width=90mm,angle=-90}}
\caption{ The angle dependence of $\phi_1 (p,q,\cos \theta)$ for different
pairs of momenta $p$ and $p$ as indicated in the figure. Note the multiplicative factors in three cases.  
\label{fig5}}
\end{figure}

\begin{figure}
\centerline{\epsfig{file=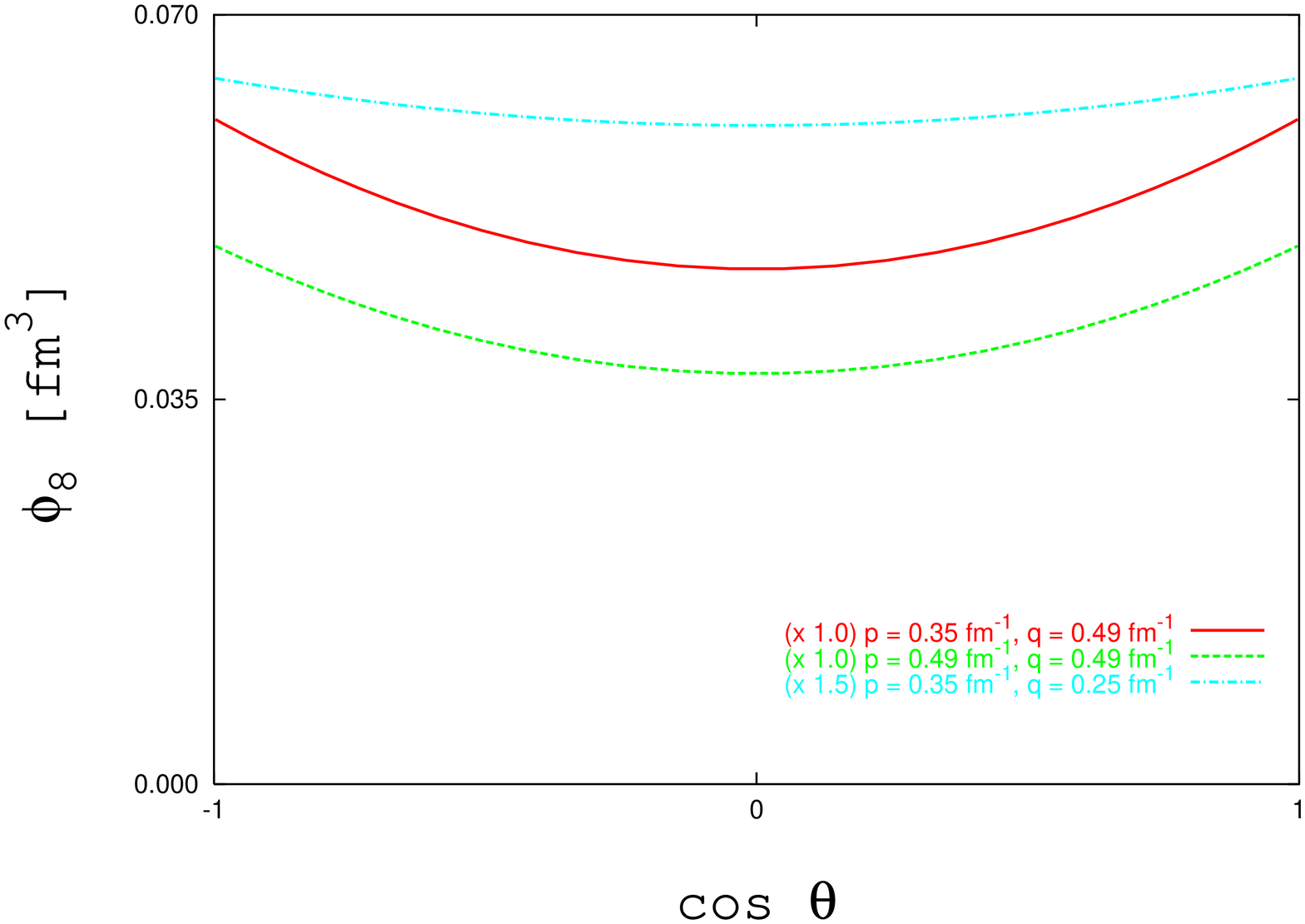,width=90mm,angle=-90}}
\caption{ The angle dependence of $\phi_8 (p,q,\cos \theta)$ for different
pairs of momenta $p$ and $p$ as indicated in the figure. Note the multiplicative factors in two cases.  
\label{fig6}}
\end{figure}

The dependence on the angle ${\bf {\hat p}}\cdot {\bf {\hat q}}$ is generally rather weak. In order
to show the angular dependence explicitly, the function
$\phi_1 (p,q,\cos \theta)$ is displayed in Fig.~\ref{fig5} for some fixed values
of $p$ and $q$ as function of $\cos \theta$. Similarly,  we show  in Fig.~\ref{fig6} the
angular dependence of $\phi_8 (p,q,\cos \theta)$. The angular dependence 
of $\phi_7 (p,q,\cos \theta)$ and $\phi_2 (p,q,\cos \theta)$ is dominantly given 
by $P_1 (\cos \theta)$, and thus not displayed.

For the calculations presented in the following, we used a wave function based on the NN force AV18 \cite{wiringa95} in conjunction with the Urbana-IX three-nucleon force \cite{pudliner97}. We show results for the $^3$He. The $\phi_i$ functions for other interactions are qualitatively similar, especially, their relative importance is not changed. Tabulated functions for several force combinations are provided by the authors \cite{urltriton}. 

In order to quantify the relative importance of the eight functions $\phi_i ({\bf p}, {\bf q})$
we consider the normalization of the 3N bound state
\begin{equation}
\langle \Psi ^{m=\frac{1}{2}} |\Psi ^{m=\frac{1}{2}} \rangle = \sum_i \int d{\bf p} \; 
d{\bf q} \; \langle \chi_i | \chi_i\rangle \phi_i^2 + 2 Re \sum_{i<j} \int d{\bf p} \; d{\bf 
q} \; \langle \chi_i | \chi_j\rangle \phi_i \phi_j
\end{equation}
using the representation given in Eq.~(\ref{eq:3.1}). The numerical evaluation 
is straightforward, and the contributions of the different products of the $\phi_i ({\bf p}, {\bf
q})$ (denoted as $N_{ij}$) to the norm are listed in Table~I. Clearly, the major contribution to the norm (91.42 \%) is
given by $\phi_1^2 ({\bf p}, {\bf q})$. The second largest contribution is already more than
one  order
of magnitude smaller and is given by $\phi_8^2 ({\bf p}, {\bf q})$. All other contributions
are even  smaller.

\section{Application}

 As example for the application of the above derived  operator form of the 3N bound
state we consider the spin dependent momentum  distribution of a
neutron inside a polarized $^3$He nucleus. This  quantity is defined as
\begin{equation}\label{eq:4.1}
N({\bf q}) = \langle \Psi^{m=\frac{1}{2}}| \delta ({\bf q} -{\bf q}_{op}) 
\frac{1}{2}(1 + \sigma_0(3)) |\Psi^{m=\frac{1}{2}}\rangle
\end{equation}
It should be noted that in the 3N c.m. system the Jacobi momentum ${\bf q}$ is the momentum of one
nucleon, here nucleon 3. Regarding  the eight operator structures displayed in Eq.~(\ref{eq:2.36})
one recognizes that only the following five different terms occur
\begin{eqnarray}
O_1 & \equiv & 1\nonumber\\
O_2 & \equiv & {\ffat \sigma}(12) \cdot {\ffat \sigma}(3)\nonumber\\
O_3 & \equiv & {\ffat \sigma}(3) \cdot {\bf A} \equiv O_3({\bf A})\nonumber\\
O_4 & \equiv & {\ffat \sigma}(12) \cdot {\bf B} \equiv O_4({\bf B})\nonumber\\
O_5 & \equiv & {\ffat \sigma}(12) \cdot {\bf C} \; {\ffat \sigma}(3) \cdot {\bf D} 
\equiv O_5({\bf C}, {\bf D}) .
\end{eqnarray}
Here the vectors {\bf A}, {\bf B}, {\bf C}, and {\bf D} represent different momentum vectors.
It is a straightforward exercise to evaluate once and for all the matrix elements
\begin{equation}
\langle O_i' O_j \rangle \equiv \langle \chi^m | O_i' O_j | \chi^m \rangle .
\end{equation}
The nonvanishing ones are listed below:
\begin{eqnarray}
\langle O_1 O_1 \rangle & = & 1\nonumber\\
\langle O_1 O_3 \rangle & = & A_0\nonumber\\
\langle O_2 O_2 \rangle & = & 3\nonumber\\
\langle O_2 O_4 \rangle & = & B_0\nonumber\\
\langle O_2 O_5 \rangle & = & {\bf C} \cdot {\bf D} + i({\bf C} \times {\bf D})_0\nonumber\\
\langle O_3'O_1 \rangle & = & A_0'\nonumber\\
\langle O_3'O_3 \rangle & = & {\bf A}' \cdot {\bf A} + i({\bf A}' \times {\bf A})_0\nonumber\\
\langle O_4'O_2 \rangle & = & B_0'\nonumber\\
\langle O_4'O_4 \rangle & = & {\bf B}' \cdot {\bf B}\nonumber\\
\langle O_4'O_5 \rangle & = & ({\bf B}' \cdot {\bf C})D_0\nonumber\\
\langle O_5'O_2 \rangle & = & {\bf C}' \cdot {\bf D}' - i({\bf C}' \times {\bf D}')_0\nonumber\\
\langle O_5'O_4 \rangle & = & ({\bf B} \cdot {\bf C}')D_0'\nonumber\\
\langle O_5'O_5 \rangle & = & ({\bf C}' \cdot {\bf C}) [{\bf D}' \cdot {\bf D} + i({\bf D}' 
\times {\bf D})_0] .
\end{eqnarray}
For the  evaluation of the specific  expectation value $N({\bf q})$ 
considered in Eq.~(\ref{eq:4.1}) we need in addition matrix elements of the form
\begin{equation}\label{eq:4.5}
\langle O_i'\sigma_0(3) O_j \rangle \equiv \langle \chi^m | O_i'\sigma_0(3) O_j | \chi^m \rangle .
\end{equation}
The resulting,  nonvanishing matrix elements are listed in Appendix~E.
With this, the specific operator  $\hat{O}$  
for evaluating $N({\bf q})$ from  Eq.~(\ref{eq:4.1}) has 
spin matrix elements given as
\begin{equation}\label{eq:4.6}
\langle O_i' \hat{O} O_j \rangle \equiv \langle O_i' \frac{1}{2}(1 + \sigma_0(3)) O_j
\rangle,
\end{equation}
and the nonvanishing matrix elements are also listed in Appendix~E. Next one expresses
the states $|\chi_i\rangle$ in terms of the 5 operators $O_i$, 
\begin{equation}
|\chi_i\rangle = \sum _j A_{ij} O_j (\Omega_{ij}) |\chi ^m\rangle ,
\end{equation}
where $\Omega_{ij}$ denotes the arguments of the operators $O_j$ which varies with $|\chi_i\rangle$.
One explicitly obtains
\begin{eqnarray}
|\chi_1\rangle &=& O_1|\chi^m\rangle \nonumber\\
|\chi_2\rangle &=& \frac{1}{\sqrt{3}} O_2 \mid \chi^m \rangle \nonumber\\
|\chi_3\rangle &=& \sqrt{\frac{3}{2}}
\frac{1}{i} \frac{ O_3({\bf p} \times {\bf q})}{pq} \mid \chi^m \rangle \nonumber\\
|\chi_4\rangle &=& \frac{1}{\sqrt{2}}
 \frac{1} {pq}  \left( i O_4({\bf p} \times {\bf q}) - O_5({\bf q}, {\bf p}) + O_5({\bf p}, {\bf q}) \right) \mid \chi^m \rangle \nonumber\\
|\chi_5\rangle &=& \frac{1}{i} \left( \frac{ O_4({\bf p} \times {\bf q}) }{pq} - \frac{i}{2}
\frac{ O_5({\bf q}, {\bf p}) - O_5({\bf p}, {\bf q}) }{pq} \right) \mid
\chi^m \rangle  \nonumber\\
|\chi_6\rangle &=& \sqrt{\frac{3}{2}} \left[ \frac{O_5({\bf p}, {\bf p})}{p^2} - \frac{1}{3} O_2 \right] \mid \chi^m \rangle \nonumber\\
|\chi_7\rangle &=& \sqrt{\frac{3}{2}} \left[ \frac{O_5({\bf q}, {\bf q})} {q^2}  -
\frac{1}{3} O_2 \right] \mid \chi^m \rangle \nonumber\\
|\chi_8\rangle &=& \frac{3}{2}  \frac{1}{\sqrt{5}} \frac{1}{pq} \left[ O_5({\bf q}, {\bf p}) 
+ O_5({\bf p}, {\bf q}) - \frac{2}{3} {\bf p} \cdot {\bf q} O_2\right] \mid \chi^m\rangle .
\end{eqnarray}
Using  the expectation values listed in Appendix~E, one can determine the matrix elements 
$\langle \chi_i|\hat{O}|\chi_j\rangle$ in a straightforward fashion. As example we give
\begin{equation}
\langle \chi_4|\hat{O}|\chi_6\rangle = \frac{\sqrt{3}}{2}\frac{1}{pq}
(\frac{i}{3}({\bf p} \times {\bf q})_0 - \frac{p_0^2}{p^2} {\bf p} \cdot {\bf q} + p_0 q_0). 
\end{equation}
Here the index 0 denotes a spherical component. The above  expression
nicely exhibits the analytic angular dependence. The final step in obtaining the
momentum distribution $N({\bf q})$ is then
to write
\begin{eqnarray}
N({\bf q}) & = & \int d{\bf p} \; \langle \Psi ^{m = \frac{1}{2}} (\hat{{\bf p}}\hat{{\bf q}})|\hat{O}|\Psi ^{m = \frac{1}{2}} (\hat{{\bf p}}\hat{{\bf q}})\rangle \nonumber\\
& = & \int d{\bf p} \; \Bigl(\sum_i \phi_i^2 \langle \chi_i |\hat{O} | \chi_i\rangle + 2 \sum_{i<j} \phi_i \phi_j Re \langle \chi_i |\hat{O} | \chi_j\rangle \Bigr) .
\end{eqnarray}

An inspection of all analytically given spin matrix elements reveals that
only four types of angular integrations occur. These are
\begin{eqnarray}
\int d\hat{{\bf p}} \; f(\hatbf{p} \cdot \hatbf{q}) 
                    \left\{ \begin{array}{l}
                    1\\
                    p_0\\
                    p_0^2\\
                    ({\bf p} \times {\bf q})_0^2
                    \end{array} \right. & &  .
\end{eqnarray}
Because of the rotational invariance around the quantization axis (z-axis)
one can choose the vector ${\bf q}$ to be in the x-z plane. Then it is most
convenient to rotate the z-axis into the  direction of ${\bf q}$ by the angle
$\theta \equiv \theta_q$. The result is
\begin{eqnarray}
\int d\hat{{\bf p}} \; f(\hatbf{p} \cdot \hatbf{q}) 
                    \left\{ \begin{array}{l}
                    1\\
                    p_0\\
                    p_0^2\\
                    ({\bf p} \times {\bf q})_0^2
                    \end{array} \right. 
 & = & 2\pi \int _{-1}^1 dx f(x)
       \left\{\begin{array}{l}
       1\\
       p \cos{\theta} x\\
       p^2 \left(\cos^2{\theta} x^2 + \frac{1}{2} \sin^2{\theta} (1 - x^2)\right)\\
       \frac{1}{2}p^2 (1 - x^2)
       \end{array}\right.   .
\end{eqnarray}
Collecting  all the results one ends with the final expression for $N({\bf q}) \equiv
N(q,\theta)$ given by
\begin{eqnarray}
N(q,\theta) & = & 2\pi \int _0^{\infty} dp \; p^2 \int _{-1}^1 dx \nonumber\\
 & & \times \Bigl\{\phi_1^2(p,q,x) + \frac{1}{3}\phi_2^2(p,q,x)\nonumber\\
 & & + \phi_2(p,q,x)\phi_6(p,q,x)\frac{1}{3\sqrt{2}} (3\cos^2{\theta} - 1) (3x^2 - 1)\nonumber\\
 & & + \phi_2(p,q,x)\phi_7(p,q,x)\frac{\sqrt{2}}{3} (3\cos^2{\theta} - 1) \nonumber\\
 & & + \phi_2(p,q,x)\phi_8(p,q,x)\frac{2}{\sqrt{15}} (3\cos^2{\theta} - 1) x \nonumber\\
 & & + \phi_3^2(p,q,x) \frac{3}{4} (1 - \cos^2{\theta}) (1 - x^2) \nonumber\\
 & & + \phi_4^2(p,q,x) \frac{1}{4} (3 + \cos^2{\theta}) (1 - x^2) \nonumber\\
 & & + \phi_4(p,q,x)\phi_5(p,q,x) \frac{1}{2\sqrt{2}} (\cos^2{\theta} - 3) (1 - x^2) \nonumber\\
 & & + \phi_4(p,q,x)\phi_6(p,q,x) \frac{\sqrt{3}}{2} (3\cos^2{\theta} - 1) (1 - x^2) x \nonumber\\
 & & + \phi_4(p,q,x)\phi_8(p,q,x) \frac{3}{2\sqrt{10}} (3\cos^2{\theta} - 1) (1 - x^2) \nonumber\\
 & & + \phi_5^2(p,q,x) \frac{1}{8} (\cos^2{\theta} + 9) (1 - x^2) \nonumber\\
 & & + \phi_5(p,q,x)\phi_6(p,q,x) \frac{1}{2}\sqrt{\frac{3}{2}} (3\cos^2{\theta} - 1) (1 - x^2) x \nonumber\\
 & & + \phi_5(p,q,x)\phi_8(p,q,x) \frac{3}{4\sqrt{5}} (3\cos^2{\theta} - 1) (1 - x^2) \nonumber\\
 & & + \phi_6^2(p,q,x) \frac{1}{12}[4 + (3\cos^2{\theta} - 1)(3x^2 - 1)] \nonumber\\
 & & + \phi_6(p,q,x)\phi_7(p,q,x) \frac{1}{6} (3\cos^2{\theta} + 1)(3x^2 - 1) \nonumber\\
 & & + \phi_6(p,q,x)\phi_8(p,q,x) \frac{1}{2\sqrt{30}} [8 + (3\cos^2{\theta} - 1)(3x^2 + 1)] x \nonumber\\
 & & + \phi_7^2(p,q,x) \frac{1}{6}(3\cos^2{\theta} + 1) \nonumber\\
 & & + \phi_7(p,q,x)\phi_8(p,q,x) \sqrt{\frac{2}{15}} (3\cos^2{\theta} + 1) x \nonumber\\
 & & + \phi_8^2(p,q,x) \frac{1}{200} [48 + (15\cos^2{\theta} - 1)(5x^2 + 3)] \Bigr\} .
\label{eq:4.12}
\end{eqnarray}
Though the angular dependence for the direction of the nucleon
momentum ${\bf q}$ in relation to the quantization axis is analytically
given, the full expression is quite lengthy. However the dependence on the
angle $\theta$ is quite simple, namely $N(q,\theta) \equiv a(q) + b(q) \cos^2 \theta$.
Moreover most of the
contributions are numerically insignificant, as illustrated in the previous section.
Therefore, we only keep  $\phi_1$, $\phi_8$, $\phi_7$, $\phi_2$. It turns out that for the specific quantity $N(q,\theta)$ only these components visibly contribute. Therefore, in this case the lengthy expression of Eq.~(\ref{eq:4.12}) shrinks  to the few leading terms
\begin{eqnarray}
N(q,\theta) & = & 2\pi \int _0^{\infty} dp \; p^2 \int _{-1}^1 dx \nonumber\\
 & & \times \Bigl\{\phi_1^2(p,q,x) + \frac{1}{3}\phi_2^2(p,q,x).\nonumber\\
 & & + \phi_2(p,q,x)\phi_7(p,q,x)\frac{\sqrt{2}}{3} (3\cos^2{\theta} - 1) \nonumber\\
 & & + \phi_2(p,q,x)\phi_8(p,q,x)\frac{2}{\sqrt{15}} (3\cos^2{\theta} - 1) x \nonumber\\
 & & + \phi_7^2(p,q,x) \frac{1}{6}(3\cos^2{\theta} + 1) \nonumber\\
 & & + \phi_7(p,q,x)\phi_8(p,q,x) \sqrt{\frac{2}{15}} (3\cos^2{\theta} + 1) x \nonumber\\
 & & + \phi_8^2(p,q,x) \frac{1}{200} [48 + (15\cos^2{\theta} - 1)(5x^2 + 3)] \Bigr\}
\label{eq:4.15}
\end{eqnarray}

\begin{figure}
\centerline{\epsfig{file=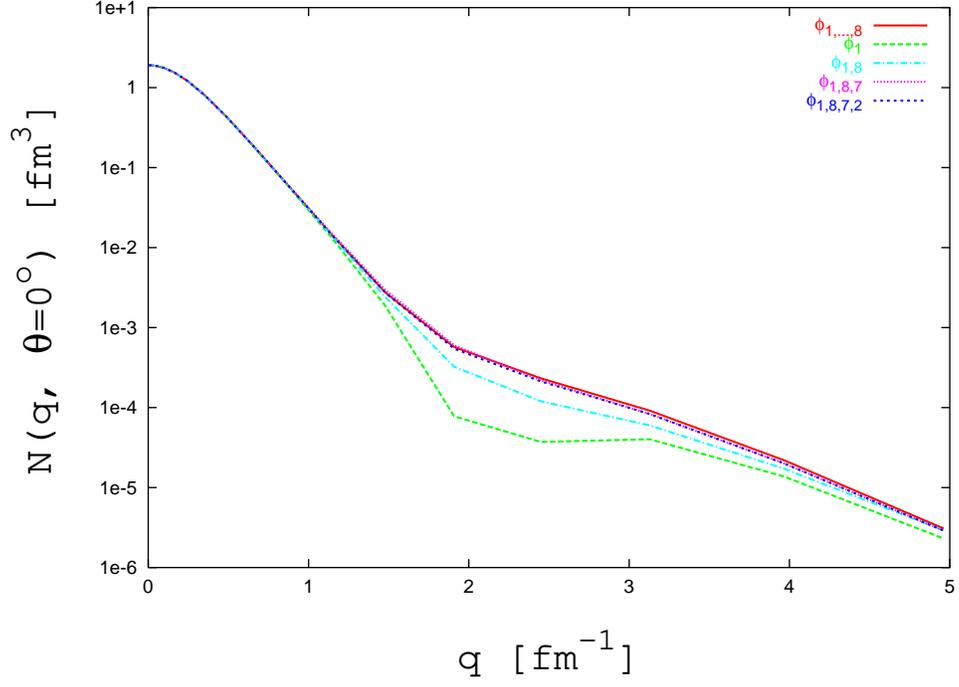,width=90mm,angle=-90}}
\caption{ The  momentum  distribution $N(q,\theta =0)$. The solid line represents the
calculation containing all terms in Eq. (\ref{eq:4.12}). The dashed line displays the result
based on the first term only, the dash-dotted the one containing in addition the
contribution of $\phi_8$. For the dotted line contributions containing $\phi_7$ are
added. The long dashed-line contains in addition contributions including $\phi_2$
and corresponds to Eq.~(\ref{eq:4.15}).
\label{fig7}}
\end{figure}

\begin{figure}
\centerline{\epsfig{file=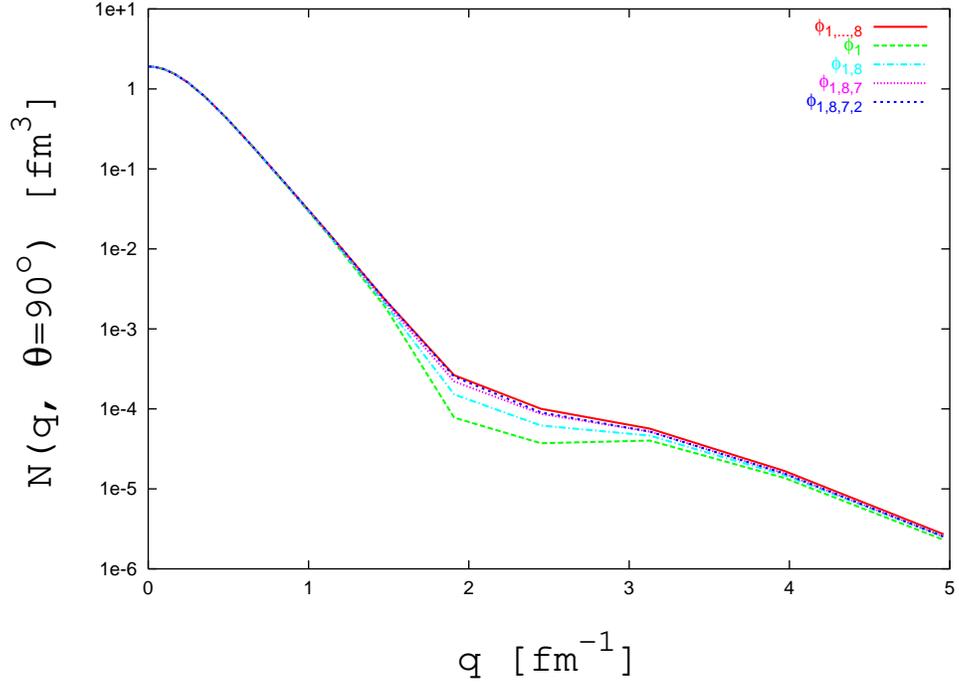,width=90mm,angle=-90}}
\caption{Same as Fig. 7, but for $N(q,\theta =90)$.
\label{fig8}}
\end{figure}

The numerical results for $N({\bf q})$ are displayed as function of $q$  in Fig.~\ref{fig7}
for the fixed angle $\theta =0$ (i.e. the nucleon momentum is parallel to the
quantization axis) and in Fig.~\ref{fig8} for $\theta =90^o$. We compare the full result given 
in Eq.~(\ref{eq:4.12}) to various truncated sums. The solid line in Figs.~\ref{fig7}
 and \ref{fig8} corresponds to the
full calculation using Eq. (\ref{eq:4.12}). The most simple approximation would be to consider
only the first term in Eq. (\ref{eq:4.12}) or Eq. (\ref{eq:4.15}), namely $\phi_1^2$, given by the 
dashed line. We see that this simple term alone already gives a good representation of 
$N(q,\theta)$ up to about $q \approx 1$~fm$^{-1}$. The dip around 2~fm$^{-1}$ is mostly 
filled in by adding the term containing $\phi_8^2$, shown as dash-dotted line. When
adding  terms containing $\phi_7$, represented by the dotted line, one is very close to
the full result. Adding the terms containing  $\phi_2$, i.e. calculating the expression
given in Eq.~(\ref{eq:4.15}), shows that all other terms in  Eq. (\ref{eq:4.12}) are insignificant. 

We expect that also for other observables the operator form of the 3N bound state will 
be useful. It should provide an easy access to the wave function without the 
need of having access to a modern triton code. 

\section{Summary}

An old idea by Gerjuoy and Schwinger \cite{gerjuoy42}  has been revived to present the
3N bound state in operator form. This form analytically  exhibits the
dominant angular and spin dependence of the wave function in form of scalar
operators formed out of momentum and spin vectors, which are  applied on a
pure spin 1/2 state. Each such operator is accompanied by a
scalar function depending on the magnitudes of the two Jacobi momenta and the
angle between them. We established the connection of this form with
the standard partial wave decomposition. This connection provided the
explicit form of the scalar functions in terms of partial wave function
components. The key point in the derivation was, to extract from an
infinite sum of partial wave expressions the operator form and the
accompanying scalar functions. The presented operator form of the
3N wave function is independent of the applied NN and 3N force.

 We illustrated the application of this new form of the
3N bound state wave function  by calculating the spin dependent single nucleon momentum
distribution in a polarized 3N  bound state. It turned out that for this
quantity only 4 parts out of the total number of 8 parts forming the  3N
bound state  were needed to achieve a sufficiently accurate
representation. Several sets of spin matrix elements depending on the
Jacobi momentum vectors, which have to be calculated only once, have
been evaluated. They will also be needed in other applications.

We expect that this operator form allows an easy access to the 3N bound
state. The 8 scalar functions carrying the specific dynamical
information  have been tabulated on a sufficiently fine grid and can be
downloaded from \cite{urltriton}. There are sets of scalar functions for various 
modern NN and 3N force combinations.

 %------------------------------------------------------------------------------

 \vfill

\begin{acknowledgments}
This work was performed in part under the
auspices of the 
U.~S.  Department of Energy under contract Nos. DE-FG02-93ER40756 (C.E.) with Ohio University, and 
DE-FC02-01ER41187 and DE-FG03-00ER41132 (A.N.). 
One of the authors (C.E.) would like to thank the Institute for Nuclear Physics at the
Forschungszentrum J\"ulich and 
the Institute for Nuclear Theory at the University
of Washington for their hospitality during some part of the work.
\end{acknowledgments}
 %\newpage

 \appendix

 \section{Relation for the coupled spherical harmonics }
 \label{appendixa}

 Here the relation of Eq.~(\ref{eq:2.10}) for the coupled spherical harmonics 
 ${\mathcal Y}^{1\mu}_{ll} ({\hat {\bf p}} {\hat {\bf q}})$ will be verified. We consider
 ${\mathcal Y}^{00}_{ll} ({\hat {\bf p}} {\hat {\bf q}}) {\mathcal Y}^{1\mu}_{11} ({\hat {\bf p}} {\hat {\bf q}})$,
 which can be rewritten by standard techniques (see e.g. \cite{glocklefb}) as
 \begin{equation} 
 {\mathcal Y}^{00}_{ll} ({\hat {\bf p}} {\hat {\bf q}}) {\mathcal Y}^{1\mu}_{11} ({\hat {\bf p}} {\hat {\bf q}})
 = \frac{3}{4 \pi} \sqrt{2l+1} \sum_{ab} (-)^{a+l} C(l1a,00) \: C(l1b,00) 
  \left\{ \begin{array} {ccc}  b \: a\:  1 \\ 1\:  1\:  l \end{array} \right\}
 {\mathcal Y}^{1 \mu} _{ab} ({\hat {\bf p}} {\hat {\bf q}}).
 \label{eq:a.1}
 \end{equation}
 The sum over $a$ and $b$ will give four terms. After inserting the explicit expression for the
 Clebsch-Gordon coefficients and the 6-j symbol, one arrives at
 \begin{equation}
 {\mathcal Y}^{00}_{ll} ({\hat {\bf p}} {\hat {\bf q}}) {\mathcal Y}^{1\mu}_{11} ({\hat {\bf p}} {\hat {\bf q}})
 = -\frac{3}{4 \pi} \frac{1}{\sqrt{2l+1}} \frac{1}{\sqrt{6}} \left( 
  -\sqrt{\frac{(l+1)(l+2)}{2l+3}} {\mathcal Y}^{1\mu}_{l+1 \, l+1} +
   \sqrt{\frac{l(l+1)}{2l-1}} {\mathcal Y}^{1\mu}_{l-1 \, l-1} \right).
 \label{eq:a.2}
 \end{equation}
 Using the relation
 \begin{equation}
 {\mathcal Y}^{00}_{ll} ({\hat {\bf p}} {\hat {\bf q}}) = (-)^l \frac{\sqrt{2l+1}}{4 \pi} P_l({\hat {\bf p}}\cdot {\hat {\bf q}}) 
 \label{eq:a.3}
 \end{equation}
 the previous Eq.~(\ref{eq:a.2}) can be rewritten as
 \begin{eqnarray}
 {\mathcal Y}^{1\mu}_{ll} ({\hat {\bf p}} {\hat {\bf q}}) & = & \sqrt{\frac{2l+1} {l(l+1)}} \nonumber\\
& & \times \left(
  (-)^{l+1} \sqrt{\frac{2}{3}} (2l-1) P_{l-1} ({\hat {\bf p}}\cdot {\hat {\bf q}})
  {\mathcal Y}^{1\mu}_{11} ({\hat {\bf p}} {\hat {\bf q}})
 + \sqrt{\frac{(l-1)(l-2)}{2l-3}} {\mathcal Y}^{1\mu}_{l-2 \, l-2} ({\hat {\bf p}} {\hat {\bf q}}) \right).  \label{eq:a.4}
 \end{eqnarray}
 This is a recursive formula and leads to the relation given in Eq.~(\ref{eq:2.10}). 
 In practice only low orbital angular momenta occur and one easily works out the lowest terms
 as
 \begin{eqnarray}
 c_1({\hat {\bf p}} {\hat {\bf q}}) &=& 1 \nonumber \\
 c_2({\hat {\bf p}} {\hat {\bf q}}) &=& -\sqrt{5} P_1 ({\hat {\bf p}}\cdot {\hat {\bf q}}) \nonumber \\
 c_3({\hat {\bf p}} {\hat {\bf q}}) &=& \frac{1}{3} \sqrt{\frac{7}{2}} \left( 5 P_2({\hat {\bf p}}\cdot {\hat {\bf q}}) +1
 \right).
 \label{eq:a.5}
 \end{eqnarray}

 \section{Verification of the relation given in Eq.~(\ref{eq:2.17})}

 The first step is to generate the spin states of Eq.~(\ref{eq:2.16}) from the states
 $\mid \chi^m \rangle$ given in Eq.~(\ref{eq:2.4}). It is easy to see that
 \begin{eqnarray}
 \mid (1 \frac{1}{2}) \frac{3}{2} \frac{3}{2} \rangle
 = \sigma(12)_{+}  \mid \chi^{m =\frac{1}{2}} \rangle,
 \label{eq:a.6}
 \end{eqnarray}
 where $\sigma_{+} (12)$
 is the spherical component of the spin operator given
 in Eq.~(\ref{eq:2.5}).

 This form can be rewritten as
 \begin{eqnarray}
 \mid (1 \frac{1}{2}) \frac{3}{2} \frac{3}{2} \rangle &=&
 \Big(\frac{2}{3}  \sigma_{+} (12)
 + \frac{1}{3} \sigma_{(+)}(12)\Big)
 \mid \chi^{m =\frac{1}{2}} \rangle
 \nonumber\\
 &=&\frac{2}{3}\Big(\sigma_{+}(12) - \frac{i}{2}
 \Big({\vec \sigma}(3) \times {\vec \sigma}(12)\Big)_{+} \Big)
 \mid \chi^{m = \frac{1}{2}} \rangle
 \label{eq:a.7}
 \end{eqnarray}
 using the spherical component of the vector product.

 Next, we express the second spin state in Eq.~(\ref{eq:2.16}) with the help of
 the lowering operator as
 \begin{eqnarray}
 \mid(1\frac{1}{2})\frac{3}{2} \frac{1}{2} \rangle &=&
 ({\mathcal S}_x - i {\mathcal S}_y)
 \mid(1\frac{1}{2})\frac{3}{2} \frac{3}{2} \rangle
 {1 \over \sqrt{3}}
 \nonumber\\
 &=& {1 \over \sqrt{6}}  \Big(\sigma_{-} (1) + \sigma_{-} (2) +
 \sigma_{-} (3) \Big)\frac{1}{2}
 \Big(\sigma_{+} (1) - \sigma_{+} (2) \Big)
 \mid \chi^{m =\frac{1}{2}} \rangle
 \nonumber\\
 &=& {1 \over \sqrt{6}} \frac{1}{2}
 \Big(2(\sigma_0(1) - \sigma_0(2) + \sigma_{-}(3)
 \Big(\sigma_{+} (1) - \sigma_{+} (2)\Big)\Big)
 \mid \chi^{m =\frac{1}{2}} \rangle
 \nonumber\\
 &=&{1 \over \sqrt{6}}\frac{1}{2}
 \Big(2(\sigma_0(1) - \sigma_0(2))
 + i \Big(({\vec \sigma}(1) -{\vec \sigma}(2)) \times
 {\vec \sigma}(3)\Big)_0 \Big)
 \mid \chi^{m =\frac{1}{2}} \rangle
 \nonumber\\
 &=&{2 \over \sqrt{6}} \Big(\sigma_0(12)-\frac{i}{2}
 ({\vec \sigma}(3) \times {\vec \sigma}(12))_0\Big)
 \mid \chi^{m =\frac{1}{2}} \rangle.
 \label{eq:a.8}
 \end{eqnarray}

\noindent
Finally, starting from
\begin{equation}
\mid(1\frac{1}{2})\frac{3}{2}- \frac{1}{2} \;\rangle \; = \;
\frac{1}{2} ({\mathcal S}_x - i
{\mathcal S}_y) \mid (1\frac{1}{2})\frac{3}{2}
\frac{1}{2} \; \rangle
\label{eq:a.9}
\end{equation}
inserting the form (B.3) and reshuffling leads to
\begin{eqnarray}
\mid(1\frac{1}{2})\frac{3}{2}- \frac{1}{2} \;\rangle \; = \;
\frac{2}{\sqrt 3}\Big( {\vec \sigma}(12)_{-} - \frac{i}{2}
({\vec \sigma}(3) \times {\vec \sigma}(12))_{-} \; \Big) \;
\mid \chi^{m =\frac{1}{2}} \rangle
\label{eq:a.10}
\end{eqnarray}

Now we use the relation (2.10) and the property
\begin{equation}
{\mathcal Y}^{1 \mu}_{11} ({\hat {\bf p}} {\hat {\bf q}}) = i
\frac{3}{4 \pi} \frac{1}{\sqrt 2}
{({\vec p} \times {\vec q})_\mu \over
\mid {\vec p} \mid \mid {\vec q} \mid}.
\label{eq:a.11}
\end{equation}
With this one arrives directly at Eq.~(\ref{eq:2.17}).

\section{Verification of the relation given in Eq.~(\ref{eq:2.20})}

With standard recoupling techniques one finds
\begin{eqnarray}
{\mathcal Y}^{00}_{ll} ({\hat {\bf p}} {\hat {\bf q}})
{\mathcal Y}^{2 m}_{11} ({\hat {\bf p}} {\hat {\bf q}}) &=&
{3 \over 4 \pi} {1 \over 2l+1} \Bigg(
\sqrt{{(l+1)(l+2)(2l+5) \over 30 (2l+3)}}
\: {\mathcal Y}^{2 m}_{l+1 l+1}
\nonumber\\
& & - \sqrt{{l(l+1) \over 5}}
({\mathcal Y}^{2 m}_{l+1,l-1} + {\mathcal Y}^{2-}_{l-1 l+1})
+ \sqrt{{l(l-1)(2l-3) \over (2l-1) \cdot 30}} \:
 {\mathcal Y}^{2 m}_{l-1,l-1}
\Bigg),
\label{eq:a.20}
\end{eqnarray}
which leads to the recursive relation
\begin{eqnarray}
{\mathcal Y}^{2m}_{ll} &=&
\sqrt{{30 (2l+1) \over l(l+1)(2l+3)}} \:
\Bigg\{ {4 \pi \over 3} (2l-1) {\mathcal Y}^{00}_{l-1 l-1}
{\mathcal Y}^{2m}_{11}
\nonumber\\
& & + \sqrt{{(l-1)l \over 5}} \: ({\mathcal Y}^{2m}_{ll-2} +
{\mathcal Y}^{2m}_{l- 2l}) -
\sqrt{{(l-1)(l-2) (2l-5)\over (2l-3) 30}} \:
{\mathcal Y}^{2m}_{l-2 l-2}\Bigg\} .
\label{eq:a.21}
\end{eqnarray}
Similarly, starting from
\begin{eqnarray}
{\mathcal Y}^{00}_{ll} ({\hat {\bf p}} {\hat {\bf q}})
Y_{2 m} ({\hat {\bf p}}) &=&
{1 \over \sqrt{4 \pi}}
\Bigg( \sqrt{3\over 2} \:
\sqrt{{(l+1)(l+2) \over (2l+1)(2l+3)}} \:
{\mathcal Y}^{2 m}_{l+2l}
\nonumber\\
& & - \sqrt{{l(l+1) \over (2l-1)(2l+3)}} \:
{\mathcal Y}^{2 m}_{ll} +  \sqrt{3\over 2}\:
\sqrt{{l(l-1) \over (2l-1)(2l+1)}} \: {\mathcal Y}^{2 m}_{l-2 l}
\Bigg),
\label{eq:a.22}
\end{eqnarray}
one finds the additional recursive relations
\begin{eqnarray}
{\mathcal Y}^{2 m}_{l+2 l}(\hat{{\bf p}}\hat{{\bf q}}) &=& \sqrt{2\over 3}\:
\sqrt{{(2l+1) (2l+3)\over (l+1)(l+2)}} \:
\Bigg\{ \sqrt{4 \pi}\: {\mathcal Y}^{00}_{ll} Y_{2m}({\hat {\bf p}})
\nonumber\\
& & + \sqrt{{l(l+1) \over (2l-1)(2l+3)}} \:
{\mathcal Y}^{2 m}_{ll} -\sqrt{3\over 2}\:
\sqrt{{l(l-1) \over (2l-1)(2l+1)}} \: {\mathcal Y}^{2 m}_{l-2,l}
\Bigg\}
\label{eq:a.23}
\end{eqnarray}
and similarly
\begin{eqnarray}
{\mathcal Y}^{2 m}_{l,l+2}(\hat{{\bf p}}\hat{{\bf q}}) &=& \sqrt{2\over 3}\:
\sqrt{{(2l+1) (2l+3)\over (l+1)(l+2)}} \:
\Bigg\{ \sqrt{4 \pi}\: {\mathcal Y}^{00}_{ll} Y_{2m}({\hat {\bf q}})
\nonumber\\
& & + \sqrt{{l(l+1) \over (2l-1)(2l+3)}} \:
{\mathcal Y}^{2 m}_{ll} -\sqrt{3\over 2}\:
\sqrt{{l(l-1) \over (2l-1)(2l+1)}} \: {\mathcal Y}^{2 m}_{l,l-2}
\Bigg\}.
\label{eq:a.24}
\end{eqnarray}
Inserting these equations into each other yields the relations given in 
Eqs.~(\ref{eq:2.20}).
For the calculation of a 3N bound state, the scalar functions $A_l$ to $E_l$ are in 
practice only needed for small values of $l$. In our context we only need odd values of $l$.
The two lowest cases are given here.

If only $l=1$ is kept, then one trivially has $A_1=1$ and $B_1=0$, and all other terms are
absent. If only $l=1$ and $l=3$ are kept, then one obtains from Eqs.~(C.4) and (C.5) 
\begin{eqnarray}
{\mathcal Y}^{2m}_{31} ({\hat {\bf p}} {\hat {\bf q}}) &=&
\sqrt{2\over 3}\:{\mathcal Y}^{2m}_{11} -
\sqrt{5\over 4 \pi} P_1 Y_{2m} ({\hat {\bf p}})
\nonumber\\
{\mathcal Y}^{2m}_{13} ({\hat {\bf p}} {\hat {\bf q}}) &=&
\sqrt{2\over 3}\:{\mathcal Y}^{2m}_{11} -
\sqrt{5\over 4 \pi} P_1 Y_{2m} ({\hat {\bf q}}).
\label{eq:a.25}
\end{eqnarray}

Furthermore, Eq.~(C.2) yields
\begin{eqnarray}
{\mathcal Y}^{2m}_{33} ({\hat {\bf p}} {\hat {\bf q}})=
\sqrt{35\over 18}
\Bigg\{
{\mathcal Y}^{2m}_{11} ({\hat {\bf p}} {\hat {\bf q}})
({5 \sqrt{5} \over 3} P_2 - \sqrt{1 \over 45}) + \sqrt{6\over 5}
({\mathcal Y}^{2m}_{13} + {\mathcal Y}^{2m}_{31}) \Bigg\}
\label{eq:a.26}
\end{eqnarray}
and after insertion of the results (C.6) we get
\begin{eqnarray}
{\mathcal Y}^{2m}_{33} ({\hat {\bf p}} {\hat {\bf q}})=
{\mathcal Y}^{2m}_{11}\sqrt{7 \over 2}
{1\over 9} (25 P_2 ({\hat {\bf p}} {\hat {\bf q}}) + 11 ) - \sqrt{35 \over 12 \pi} 
P_1({\hat {\bf p}} {\hat {\bf q}}) \Big(Y_{2m} ({\hat {\bf p}}) + Y_{2m} ({\hat {\bf q}})\Big).
\label{eq:a.27}
\end{eqnarray}

\noindent
Thus, one obtains in the end the coefficients given in Eq.~(\ref{eq:2.30}).

\section{The Operators of  Eq.~(\ref{eq:2.27})}

The starting points for the derivation
 are Eqs.~(\ref{eq:2.18}) and (\ref{eq:2.23}). According to 
Eqs.~(\ref{eq:a.7}), (\ref{eq:a.8}), and (\ref{eq:a.10}) these lead to 
\begin{eqnarray}
\mid \, ^4 D^{m=\frac{1}{2}}_{\frac{1}{2}}\rangle 
&=&  - {2 \over 3} \sqrt{1 \over 10}
 \Big( \sigma(12) - {i \over 2}(\sigma(3) \times \sigma(12))\Big)
_+ \mid \, \chi^{m=\frac{1}{2}}\rangle \nonumber \\
& & \times \sum_{lodd} \Big( {\mathcal Y}^{2-1}_{11} ({\hat {\bf p}} {\hat {\bf q}})
X_l + Y_{2-1} ({\hat {\bf p}}) V_l + Y_{2-1}({\hat {\bf q}}) W_l\Big) \nonumber\\
& & + \sqrt{1 \over 5} {2 \over \sqrt{6}} \Big(\sigma(12)
- {i \over 2} \sigma(3) \times \sigma(12)\Big)_0
\mid \, \chi^{m=\frac{1}{2}}\rangle \nonumber\\
& & \times \sum_{lodd} \Big( {\mathcal Y}^{20}_{11} ({\hat {\bf p}} {\hat {\bf q}}) X_l +   
 Y_{20}({\hat {\bf p}})V_l + Y_{20}({\hat {\bf q}})W_l\Big)  \nonumber\\
& & - \sqrt{3 \over 10}  {2 \over \sqrt{3}}\Big(\sigma
(12) - {i \over 2}(\sigma(3) \times \sigma(12))_{-}
\mid \chi^{m=\frac{1}{2}}\rangle \nonumber\\
& & \times \sum_{lodd}\Big({\mathcal Y}^{21}_{11} ({\hat {\bf p}} {\hat {\bf q}})
X_l + Y_{21}({\hat {\bf p}})V_l + Y_{21}({\hat {\bf q}}) W_l \Big)
\nonumber\\
& & + \sqrt{2 \over 5}  \sqrt{1 \over 2} \sigma_{-}(3) \sigma_{-}(12)\mid
\chi^{m=\frac{1}{2}}\rangle
 \sum_{lodd}\Big({\mathcal Y}^{22}_{11} ({\hat {\bf p}} {\hat {\bf q}})
X_l + Y_{22}({\hat {\bf p}})V_l + Y_{22}({\hat {\bf q}}) W_l \Big) .
\label{eq:a.30}
\end{eqnarray}
As an example let us regard the terms in $Y_{2 \mu}({\hat {\bf p}})$, which have the well known
representation
in terms of spherical components of ${\bf p}$:
\begin{eqnarray}
Y_{2-1} ({\hat {\bf p}}) &=& {1\over 2 {\bf p}^2} \sqrt{15\over \pi}
p_0 p_{-}
\nonumber\\
Y_{20}({\hat {\bf p}}) &=& {1\over 2 {\bf p}^2} \sqrt{5\over \pi}
\left( p_0^2 +p_{+} p_{-} \right)
\nonumber\\
Y_{21}({\hat {\bf p}}) &=& {1\over 2 {\bf p}^2} \sqrt{15\over \pi}
p_0p_{+}
\nonumber\\
Y_{22}({\hat {\bf p}}) &=& {1\over 2 {\bf p}^2} \sqrt{15\over 2\pi}
(p_{+})^2 .
\end{eqnarray}

For the convenience of the reader we also provide the relations
\begin{eqnarray}
{\mathcal Y}^{2-1}_{11} ({\hat {\bf p}} {\hat {\bf q}}) &=& {1\over \sqrt{2}}
{3\over 4\pi} {1 \over \mid {\bf p} \mid \mid {\bf q}\mid}
\left( p_0 q_{-} + p_{-} q_0 \right)
\nonumber\\
{\mathcal Y}^{20}_{11}({\hat {\bf p}} {\hat {\bf q}}) &=& {3\over 4\pi}
{1 \over \mid {\bf p} \mid \mid {\bf q}\mid} \left( {1\over \sqrt{6}}
(p_{+} q_{-} + p_{-} q_{+})+ \sqrt{{2\over 3}} p_0 q_0 \right)
\nonumber\\
{\mathcal Y}^{21}_{11}({\hat {\bf p}} {\hat {\bf q}}) &=& {1\over \sqrt{2}}
{3\over 4\pi} {1 \over \mid {\bf p}\mid  \mid {\bf q}\mid} 
\left( p_{+} q_{0} + p_{0} q_{+} \right)
\nonumber\\
{\mathcal Y}^{22}_{11}({\hat {\bf p}} {\hat {\bf q}}) &=& {3\over 4\pi}
{p_{+} q_{+} \over \mid {\bf p} \mid \mid {\bf q} \mid} .
\end{eqnarray}

\noindent
If one now inserts (D.2) into (D.1) and looks only into the terms with $V_l$, one can
easily combine the expressions to
\begin{equation}
\mid \, 4D^m_{\frac{1}{2}}\rangle \Big|_{V_l} =
 \frac{1}{2} {\sqrt{3\over 2 \pi}}
\sum_{lodd}V_l
\left[ {{\bf \sigma}(12) \cdot {\bf p} \; {\bf \sigma} (3) \cdot {\bf p}
\over {\bf p}^2} - \frac{1}{3}{\bf \sigma}(12) \cdot{\bf \sigma}(3)
\right] \mid \chi^m \rangle
\end{equation}
The term in $X_l$ is somewhat more tedious.

\section{The Nonvanishing Matrixelements of Eqs.~(\ref{eq:4.5}) and (\ref{eq:4.6})}

The nonvanishing matrix elements of Eq.~(\ref{eq:4.5}) are given by

\begin{eqnarray}
\langle O_1 \sigma_0(3) O_1 \rangle & = & 1 \nonumber\\
\langle O_1 \sigma_0(3) O_3 \rangle & = & A_0 \nonumber\\
\langle O_2 \sigma_0(3) O_2 \rangle & = & -1 \nonumber\\
\langle O_2 \sigma_0(3) O_4 \rangle & = & B_0 \nonumber\\
\langle O_2 \sigma_0(3) O_5 \rangle & = & 2C_0D_0 - {\bf C} \cdot {\bf D} - i({\bf C} \times {\bf D})_0 \nonumber\\
\langle O_3'\sigma_0(3) O_1 \rangle & = & A_0' \nonumber\\
\langle O_3'\sigma_0(3) O_3 \rangle & = & 2A_0'A_0 - {\bf A}' \cdot {\bf A} - i({\bf A}' \times {\bf A})_0 \nonumber\\
\langle O_4'\sigma_0(3) O_2 \rangle & = & B_0' \nonumber\\
\langle O_4'\sigma_0(3) O_4 \rangle & = & {\bf B}' \cdot {\bf B} \nonumber\\
\langle O_4'\sigma_0(3) O_5 \rangle & = & ({\bf B}' \cdot {\bf C}) D_0 \nonumber\\
\langle O_5'\sigma_0(3) O_2 \rangle & = & 2C_0'D_0' - {\bf C}' \cdot {\bf D}' + i({\bf C}' \times {\bf D}')_0 \nonumber\\
\langle O_5'\sigma_0(3) O_4 \rangle & = & ({\bf B} \cdot {\bf C}') D_0' \nonumber\\
\langle O_5'\sigma_0(3) O_5 \rangle & = & ({\bf C}' \cdot {\bf C}) 
[2D_0'D_0 - {\bf D}' \cdot {\bf D} - i({\bf D}' \times {\bf D})_0 ].
\end{eqnarray}

The nonvanishing matrix elements of Eq.~(\ref{eq:4.6}) are given by

\begin{eqnarray}
\langle O_1 \hat{O} O_1 \rangle & = & 1 \nonumber\\
\langle O_1 \hat{O} O_3 \rangle & = & A_0 \nonumber\\
\langle O_2 \hat{O} O_2 \rangle & = & 1 \nonumber\\
\langle O_2 \hat{O} O_4 \rangle & = & B_0 \nonumber\\
\langle O_2 \hat{O} O_5 \rangle & = & C_0D_0 \nonumber\\
\langle O_3'\hat{O} O_1 \rangle & = & A_0' \nonumber\\
\langle O_3'\hat{O} O_3 \rangle & = & A_0'A_0 \nonumber\\
\langle O_4'\hat{O} O_2 \rangle & = & B_0' \nonumber\\
\langle O_4'\hat{O} O_4 \rangle & = & {\bf B}' \cdot {\bf B} \nonumber\\
\langle O_4'\hat{O} O_5 \rangle & = & ({\bf B}' \cdot {\bf C}) D_0 \nonumber\\
\langle O_5'\hat{O} O_2 \rangle & = & C_0'D_0' \nonumber\\
\langle O_5'\hat{O} O_4 \rangle & = & ({\bf B} \cdot {\bf C}') D_0' \nonumber\\
\langle O_5'\hat{O} O_5 \rangle & = & ({\bf C}' \cdot {\bf C}) D_0'D_0 . 
\end{eqnarray}

%------------------------------------------------------------------------------
%\pagebreak

\bibliography{lit150104}

%\pagebreak
%%%%%%%%%%%%%%%%%%%%%%%%%%%%%%%%%%%%%%%%%%%%%%%%%%%%%%

\begin{table}[!ht]
\caption{\label{table1}The contributions 
$N_{ij}$ from $\phi_i({\bf p},{\bf q}) \phi_j({\bf p},{\bf q})$ to the total
 normalization of the 3N state.}
\begin{tabular}{|c||c|c|c|c|c|c|c|c|c|c|c|}
\hline
 $ij$ &    $11$ & $22$ & $33$ & $44$ & $55$ & $66$ & $67$ & $68$ & $77$ & $78$ &  $88$ \\  
\hline
$N_{ij}$ [\%]& 91.42 & 0.76 & 0.02 & 0.02 & 0.02& 0.13 & 0.11& 0.25& 0.98&  2.44& 4.35 \\ 
\hline 
\end{tabular}
\end{table}

%\pagebreak
%%%%%%%%%%%%%%%%%%%%%%%%%%%%%%%%%%%%%%%%%%%%%%%%%%%%%%

%\noindent

%\noindent

%\noindent

%\noindent

%\noindent

%\noindent

%\noindent

%\noindent
  
%%%%%%%%%%%%%%%%%%%%%%%%%%%%%%%%%%%%%%%%

%\pagebreak

%\input{psfig}

%\centerline{Fig. 1}
%\centerline{\psfig{file=phi1_0he3.ps,width=150mm,angle=-90}}
%\includegraphics*[width=5.0cm,angle=-90]{phi1_0.ps}

%\newpage
%\centerline{Fig. 2}
%\centerline{\psfig{file=phi8_0he3.ps,width=150mm,angle=-90}}

%\newpage
%\centerline{Fig. 3}
%\centerline{\psfig{file=phi7_0he3.ps,width=150mm,angle=-90}}

%\newpage
%\centerline{Fig. 4}
%\centerline{\psfig{file=phi2_0he3.ps,width=150mm,angle=-90}}

%\newpage
%\centerline{Fig. 5}
%\centerline{\psfig{file=phi1he3.angle.ps,width=150mm,angle=-90}}

%\newpage
%\centerline{Fig. 6}
%\centerline{\psfig{file=phi8he3.angle.ps,width=150mm,angle=-90}}

%\newpage
%\centerline{Fig. 7}
%\centerline{\psfig{file=probconv0he3.ps,width=150mm,angle=-90}}

%\newpage
%\centerline{Fig. 8}
%\centerline{\psfig{file=probconv90he3.ps,width=150mm,angle=-90}}

\end{document}